\documentclass{comjnl}

\usepackage{amsmath}
\usepackage{algorithm}
\usepackage{algorithmic}  
\usepackage[algo2e,ruled]{algorithm2e} 
\usepackage{graphicx}
\usepackage{float}
\usepackage{enumerate}
\usepackage{array}
\usepackage{amssymb}
\usepackage{color}
\usepackage{soul}
\usepackage{amsmath}
\usepackage{caption,subcaption}

\newcommand{\resetlinenumber}{\setcounter{AlgoLine}{0}}

\begin{document}

\title[Individualized Time-Series Segmentation]{Individualized Time-Series Segmentation for Mining Mobile Phone User Behavior}
\author{Iqbal H. Sarker$^{1,*}$, Alan Colman$^1$, Muhammad Ashad Kabir$^2$ and Jun Han$^1$}
\affiliation{$^1$ Department of Computer Science and Software Engineering, \\ 	  School of Software and Electrical Engineering, \\ 
		Swinburne University of Technology, \\ Melbourne, Australia.\\
		$^2$ School of Computing and Mathematics, \\
		Charles Sturt University, NSW, Australia.\\
		} \email{{\{*Corresponding author: msarker@swin.edu.au \}}}

\shortauthors{Sarker et al.}

\received{00 January 2009}
\revised{00 Month 2009}

\keywords{Mobile data mining; machine learning; clustering; time-series; segmentation; behavior prediction; temporal rule discovery; mobile phone user; context; personalization. \\ (Published in The Computer Journal, Oxford Unviersity, Uk, 61(3), 2017.)}

\begin{abstract}
Mobile phones can record individual's daily behavioral data as a time-series. In this paper, we present an effective time-series segmentation technique that extracts \textit{optimal time segments} of individual's similar behavioral characteristics utilizing their mobile phone data. One of the determinants of an individual's behavior is the various activities undertaken at various times-of-the-day and days-of-the-week. In many cases, such behavior will follow temporal patterns. Currently, researchers use either \textit{equal} or \textit{unequal} interval-based segmentation of time for mining mobile phone users' behavior. Most of them take into account \textit{static} temporal coverage of 24-hours-a-day and few of them take into account the \textit{number of incidences} in time-series data. However, such segmentations do not necessarily map to the patterns of individual user activity and subsequent behavior because of not taking into account the \textit{diverse behaviors} of individuals over time-of-the-week. Therefore, we propose a \textit{behavior-oriented time segmentation (BOTS)} technique that takes into account not only the temporal coverage of the week but also the \textit{number of incidences} of diverse behaviors dynamically for producing similar behavioral time segments over the week utilizing time-series data. Experiments on the real mobile phone datasets show that our proposed segmentation technique better captures the user's \textit{dominant behavior} at various times-of-the-day and days-of-the-week enabling the generation of high confidence \textit{temporal rules} in order to mine individual mobile phone users' behavior.
\end{abstract}

\maketitle

\section{Introduction}
\label{Introduction}
Now-a-days, mobile phones have become part of our life. The number of mobile cellular subscriptions is almost equal to the number of people on the planet \cite{pejovic2014interruptme}. The phones are, for most of the day, with their owners as they go through their daily routines \cite{pejovic2014interruptme}. People use mobile phones for various activities such as voice communication, Internet browsing, apps using, e-mail, online social network, instant messaging etc. \cite{pejovic2014interruptme}. Their ability to log such activities offers the potential to understand the behavior of individual mobile phone users. In recent years, researchers have used various types of mobile phone data such as phone call logs \cite{phithakkitnukoon2011behavior}, app usages logs \cite{srinivasan2014mobileminer}, mobile phone notification logs \cite{mehrotra2016prefminer}, web logs \cite{halvey2005time}, context logs \cite{zhu2014mining} to mine individual mobile user's behavior for different purposes. For instance, in order to build an automated call firewall or call reminder systems, phone call log is used to predict users' phone call behavior \cite{phithakkitnukoon2011behavior}.

Time is the most important factor that impacts user behavior in a mobile-Internet portal \cite{halvey2006time}. As individual's behaviors vary over time, the devices record the exact time (e.g., 2015-04-25 08:35:55) of all diverse behaviors with mobile phones (the ``time series data") of the users. However, human understanding of time is not precise, unlike digital systems. There is always a time interval for routine behaviors, even if only a small interval, e.g., five minutes. For instance, a user regularly makes a phone call to her mother in the evening. It is unlikely that she will call her mother everyday exactly at 6:00PM; she could call one day at 6:13PM and another day at 5:51PM. Therefore, in the time prediction of user behavior, exact time is not very informative. According to \cite{farrahi2014probabilistic} time-based effective behavior modeling is an open problem. In this paper, we focus on mining mobile user behavior based on time by extracting \textit{similar behavioral time segments} for various days-of-the-week enabling the generation of high confidence \textit{temporal rules} utilizing time-series data.

To evaluate time as a condition in a high confidence rule, time must be segmented into meaningful categories that serve as a proxy for identifying user's diverse behaviors. To mine mobile user behavior for different purposes, researchers use \textit{equal} or \textit{unequal} interval-based segmentation of time that includes either \textit{large interval} or \textit{small interval} without taking into account individual's behavioral patterns. For instance, a number of researchers \cite{mehrotra2016prefminer, song2013exploring, xu2013preference, mukherji2014adding} use \textit{large interval} based segmentation (e.g., morning[6:00AM-12:00PM]) in order to mine mobile user behavior. However, such large segments are not suitable for the production of meaningful behavioral rules of individuals. Let us consider a phone call example. Say, on Monday a user attends on a regular meeting from 8:00AM to 8:30AM while another user attends class form 10:00AM to 11:30AM. Both users reject the incoming call while in meeting or class. At other times in morning the users may typically answer the incoming calls. By using static large segments (e.g., morning) these logged call response behaviors would not be generalizable to a meaningful rule because of not able to differentiate individual's such diverse behaviors in morning.

On the other hand, a number of researchers use \textit{small interval} based segmentation (e.g., 15 minutes) \cite{ozer2016predicting, phithakkitnukoon2010activity, do2014and} instead of the above large categories by taking into account the frequent variations of individual's behaviors. However, in many cases meaningful rules will not be found using these small interval time segments. For example, if the time interval is very small, there may not be enough behavioral data instances in each segment to determine the \textit{dominant behavior} based on multiple observations, or there may be no data at all for that segment. Creating behavioral rules based on observations with so little ``support'' (data instances) is unlikely to be effective. In general, by increasing the time interval we would expect more data instances (greater \textit{support}) but also greater behavioral variations to be observed - thus it masks the actual \textit{dominant} behavior. Since each individual's behavior is different, such segments are not suitable for capturing the actual behaviors of mobile phone users. Therefore, for producing effective temporal rules, individual's \textit{behavior-oriented} time segments need to be discovered that reflect the logged behavior of an individual mobile phone user.

In addition to the time of a day, the specific \textit{day-of-the-week} needs to be considered to get pertinent rules. For many users their daily schedule differs from day-to-day. For instance, a user has a meeting on every Friday[2:00PM-3:00PM] and rejects the incoming calls during that time period, but on other days he is available at that time and answers the incoming calls as usual. If we don't differentiate user behaviors between days-of-the-week, the other days' different behaviors will mask the \textit{dominant behavior} on Friday, and we would thereby falsely conclude that a reject behavioral rule at that time period on Friday has no significance. 

To address the above problems, we propose an approach that analyses an individual's mobile phone time-series data and discovers the \textit{behavior-oriented time segments} in order to mine an individual's behavior. An effective segmentation of time will produce \textit{high confidence rules} that capture dominant behavior over as much of the week as possible. To produce rules, we use association rule learner \cite{agrawal1994fast} rather than using classification rule learner. According to \cite{freitas2000understanding}, classification learners cannot ensure that a discovered classification rule will have a high predictive accuracy. In contrast, association rule learning is a well-defined, deterministic task that discovers the rule sets having confidence greater than a preferred threshold. The setting of this threshold for creating rules will vary according to an \textit{individual's preference} as to how interventionist they want the agent to be. 

Let's consider the phone call-handling agent as an example. One person may want the agent to reject calls where in the past he/she has rejected calls more than, say, 80\% of the time - that is, at a threshold of 80\%. Another individual, on the other hand, may only want the agent to intervene if he/she has rejected calls in, say, 95\% of past instances. However, the traditional metrics `confidence' and `support' of association rule learner \cite{agrawal1994fast} are not sufficient to identify the \textit{optimal segments} for producing effective temporal rules because of not taking into account the volatility of an individual's behavior over time. Therefore, to establish the \textit{optimal segmentation}, we propose a \textit{metric} `applicability' (in addition to traditional `support' and `confidence') that measures the applicability of rules generated by that segmentation. \textit{Applicability} is a descriptive statistic that measures how much of the week is covered by rules (the ``temporal coverage") and takes into account the data instances in each time segment (the ``support"), for a particular confidence threshold.

In our technique, we follow \textit{bottom-up} processing of individual's mobile phone data to achieve our goal. We initially divide each day of the week into relatively small time slices using a small \textit{base period} and identify the dominant behavior for each slice. After that, we dynamically aggregate adjacent slices with the same dominant characteristics to get larger segments of similar behavior. These larger time segments will have more support and are then used as the basis for mining rules pertinent to the individuals. The \textit{applicability} for that segmentation is then measured. As we have no prior knowledge about individual's behavioral patterns, we then iteratively increment the base period and compare the corresponding applicability over each iteration in order to identify the optimal base period. The time segmentation that yields the maximum applicability establishes the \textit{optimal time segmentation}, and the corresponding base period is \textit{optimal base period} that captures the unique behavioral patterns of individuals. Finally, the generated rules for the optimal segmentation will be the output for the users. As the behaviors of different individuals are not identical in the real word, such segmentation may differ from user-to-user according to their behavioral patterns over time-of-the-week. 

The contributions are summarized as follows.
\begin{itemize}
	\item We propose a \textit{metric `applicability'} that takes into account both the temporal coverage and support of a segment, in order to identify the \textit{optimal} time-series segmentation.
	\item We propose a \textit{behavior-oriented time segmentation} technique for mining individualized time-dependent behavioral rules of mobile phone users utilizing their phone log data.
	\item Our experiments on real datasets show that this segmentation technique is more effective than existing techniques for mining user behavior when applied to mobile phone data.   
\end{itemize}

This paper significantly revises and extends \cite{sarker2016behavior} by elaborating the BOTS technique in several directions: (i) defining and formulating the problem statement clearly in terms of mathematical notation; (ii) taking into account the impact on day-wise behavioral variations of individuals for effective segmentation; (iii) introducing an efficient way to identify the optimal similar behavioral segmentation; (iv) a range of experiments have been conducted on the real-world mobile phone datasets (Massachusetts Institute of Technology (MIT) Reality Mining \cite{eagle2006reality}); (v) additional evaluation measurements have been taken into account to evaluate the segmentation quality and corresponding temporal rules as well; (vi) showing the effect on each parameter used in the technique by experiments; (vii) extending more recent related works and summarizing a number of real-world applications.

The rest of the paper is organized as follows. Section \ref{Related Work} provides a brief review of related work. In section \ref{Problem Statement}, we define and formulate the problem addressing in this paper. In Section \ref{Our Approach}, we present our behavior-oriented time segmentation approach for discovering temporal rules of individuals. We report the experimental results in Section \ref{Experiments}. Some key observations of our technique are summarized in Section \ref{Discussion}. A number of real-world applications of behavior-oriented segments are mentioned in Section \ref{Applications of BOTS} and finally Section \ref{Conclusion and Future Work} concludes this paper and highlights future work.

\section{Related Work}
\label{Related Work}
In recent years, variety of time series segmentations for mining mobile phone user behavior have been used in various purposes. However, such segmentations are not individualized \textit{behavior-oriented}. There are mainly two types of time intervals: one is \textit{equal} and another one is \textit{unequal} that are used in segmentation approaches \cite{zhang2015time}. Based on these two intervals, in this section we review different time segmentation approaches that are used in various purposes.

\subsection{Equal interval-based segmentation}
A number of authors have used equal interval-based segmentation in their applications, such as Song et al. \cite{song2013exploring} present a log based study on users' search behavior to improve search relevance by dividing 24-hours-a-day into three equal time segments, e.g., morning [7:00-12:00], afternoon [13:00-18:00] and evening [19:00-24:00]. Mukherji et al. \cite{mukherji2014adding} take into account four time segments, i.e., morning [6:00-12:00], afternoon [12:00-18:00], evening [18:00-24:00] and night [0:00-6:00]. In \cite{paireekreng2009time}, Paireekreng et al. have proposed a personalization mobile game recommendation system using time-of-the-day divided into 4 periods - morning, afternoon, evening and night respectively. To understand the variation in variety seeking over different time windows Jayarajah et al. \cite{jayarajah2014exploring} use morning [6:00-11:59], day [12:00-17:59], evening [18:00-23:59], overnight [0:00-5:59]. Do et al. \cite{do2010their} use night [0:00AM-6:00AM], morning [6:00AM-12:00PM], afternoon [12:00PM-6:00PM], and evening [6:00PM-0:00AM] to understanding how the user behavior changes with respect to the time of the day in their application model. Rawassizadeh et al. \cite{rawassizadeh2016scalable} propose a scalable approach for daily behavioral pattern mining from multiple sensor data using three temporal segments [0:00-7:59], [8:00-15:59] and [16:00-23:59].

Besides such segmentations, a number of researchers use a single parameter `time interval length' to define varying length time intervals for time segmentation. As a result, each day is divided into a predefined number of equivalent length time intervals. For instance, Ozer et al. \cite{ozer2016predicting} propose an approach to predict the location and time of mobile phone users by using sequential pattern mining techniques. In their approach, they use 15 minutes as a time interval length for segmentation. In \cite{do2014and}, Do et al. present a framework for predicting where users will go and which app they will use in the next by exploiting the rich contextual information from smartphone sensors. In their framework, they use 48 equal streams for 24-hours-time-of-the-day. In \cite{farrahi2008did}, Farrahi et al. use temporal data to discover daily routines from large-scale mobile phone data. They divide each day into 30-minute time-slots resulting in 48 blocks per day. In \cite{karatzoglou2012climbing}, Karatzoglou et al. use the time of the day in blocks of 2-hours in their mobile app recommendation system. Phithakkitnukoon et al. \cite{phithakkitnukoon2010activity} use 3-hours interval for time segmentation in their study to identify human daily activity patterns using mobile phone data. 

However, the above segmentations do not take into account the behavioral evidence that differs from user-to-user over time-of-the-week. As a result, these static segmentations are not suitable for producing high confidence temporal rules of individuals.

\subsection{Unequal interval-based segmentation}
A number of authors have used unequal interval-based segmentation in their applications, such as Xu et al. \cite{xu2013preference} have presented a prediction framework for smartphone app usages by incorporating three important everyday factors (context, community behavior and user preferences) that influence user app usages behavior. In their approach, they use morning (beginning at 6:00AM and ending at noon), afternoon (ending at 6:00PM), night (all remaining hours) for time segmentation. In \cite{mehrotra2016prefminer}, Mehrotra et al. propose a novel interruptibility management solution that learns users’ preferences for receiving mobile notifications based on automatic extraction of rules by mining their interaction with mobile phones. For segmentation, they use four time slots – morning [6:00-12:00], afternoon [12:00-16:00], evening [16:00-20:00] and night [20:00-24:00 and 0:00-6:00]. Zhu et al. \cite{zhu2014mining} use five static time segments in a day and predefined as morning [7:00-11:00], noon [11:00-14:00], afternoon [14:00-18:00] and so on in their recommendation system. To describe the feelings, ideas, opinions, and emotions of each user, Oulasvirta et al. \cite{oulasvirta2012habits} use five time slots (morning, forenoon, afternoon, evening, and night) as temporal context. In \cite{yu2012towards}, Yu et al. investigate how to exploit user context logs for personalized context-aware recommendation by mining CCPs through topic models. In their system, they use morning [7:00-11:00], noon [11:00-14:00], afternoon [14:00-18:00], evening [18:00-21:00], and night [21:00-Next day 7:00] for time segmentation.

In addition to the above time segments, a number of authors \cite{naboulsi2014classifying, dashdorj2013semantic, shin2009context} introduce early morning, late morning, midnight and so on statically. For instance, in \cite{shin2012understanding} Shin et al. propose a new context model for app prediction, which collects a wide range of contextual information in a smartphone and makes personalized app predictions based on naive Bayes model. In their model, they categorize time into early morning, morning, afternoon, evening, night for weekday and weekend. In \cite{farrahi2010probabilistic}, Farrahi et al. divide each day into 8 coarse-grain time slots as follows: [0:00AM-7:00AM], [7:00AM-9:00AM], [9:00AM-11:00AM], [11:00AM-2:00PM], [2:00PM-5:00PM], [5:00PM-7:00PM], [7:00PM-9:00PM] and  [9:00PM-12:00AM]. These time slots were chosen to capture common events in daily life, such as lunch time, dinner time, or morning and afternoon work times. Such segmentations are also used in various applications such as managing mobile intelligent interruption management system \cite{zulkernain2010mobile}, making app prefetch practical on mobile phones \cite{parate2013practical}, mining frequent co-occurrence patterns on the mobile phones \cite{srinivasan2014mobileminer}, mining mobile user habits \cite{ma2012habit, cao2010effective}. However, these static segmentations do not take into account the behavioral evidence that differs from user-to-user over time-of-the-week.

To identify dynamic segmentation using mobile phone data, Das et al. \cite{das1998rule} propose a cluster-based technique in order to discover rules from time-series. However, the problem is that the number of clusters has to be known in advance that is difficult to assume for an individual. In order to predict mobile user navigation patterns, Halvey et al. \cite{halvey2005time} have presented a multi-thresholds based method for segmenting time-series log data. However, it is very difficult to choose these thresholds that are used to identify the lower and upper boundary of a segment because of having no prior knowledge about user activities. Besides these, GA based \cite{lu2011mining, kandasamy2015modified}, sliding window based \cite{hartono2014extracting}, shape based  \cite{zhang2015time, shokoohi2015discovery} segmentation have been proposed for different purposes. These segmentations are based on the total number of activity occurrences of the user at each time point. However, these are not behavior-oriented segmentations as they do not take into account diverse behaviors of individuals, in which we are interested in.

A number of authors analyze user diverse behaviors in different time periods utilizing mobile phone data. For instance, Phithakkitnukoon et al. \cite{phithakkitnukoon2011behavior} design a behavior-based adaptive call prediction utilizing mobile phone data. In \cite{jang2015combination}, Jang et al. have shown that different users app usages behavior varies over time in a day utilizing mobile phone data. In \cite{henze2011release}, Henze et al. utilize mobile phone data in order to find the best time to deploy apps. To identify the suitable time period of active apps, Xu et al. \cite{xu2011identifying} utilize mobile phone data. Bohmer et al. \cite{bohmer2011falling} identify the peak time of average app usages based on user behavior. These approaches take into account the scanning over each hour time slot of the day (e.g., [1:00PM-2:00PM]), for capturing user behaviors and identify a particular predefined segment for their own purposes. However, such approaches do not take into account the dynamic optimal segmentation according to individual's behavior. 
		
Unlike these works, we identify the number of optimal segments dynamically without any prior knowledge by analyzing individual's similar behavioral patterns and extract a set of effective time segments with associated days for producing high confidence temporal behavior rules of individuals. 

\section{Problem Statement}
\label{Problem Statement}
Let $Db$ be a mobile phone dataset with an attribute $A$ that represents temporal information in time-series and $|Db|$ denotes the number of records in $Db$, where each record has an identifier $T_{id}$. Let $BHs = \{BH_1,BH_2, ..., BH_n\}$ be a list of behaviors with mobile phones of an individual user and $n$ is the total number of behavior classes. A specific value of the time-series attribute $A$ and behavior class $BH_j$ are denoted by lower-case letters $a_i$ and $bh_j$, respectively.

\textbf{Definition (Time-Series).} A time-series $T_{series}$ is a sequence of data points ordered in time such that $T_{series} = (t_1, t_2,..., t_m)$, where $t_1, t_2,...,t_m$ are individual observations, each of which contains real-value data and $m$ is the number of observations in a time series.

\textbf{Definition (User Behavior).} The behaviors $BHs$ of an individual user $U$ represents different activities or usages habits with mobile phone that are logged by the device in time-series. Let $Hs = \{H_1, H_2, ,..., H_m\}$ be a list of usages habits of a mobile phone user and $m$ is the number of observations in a time series, then $BHs = distinct \{H_1, H_2, ..., H_m\}$.

\textbf{Definition (Behavioral Transaction).} A behavioral transaction is a set of raw data such as $BT =(T_{id}, T_t, T_a, T_o)$, where $T_{id}$ is an identifier of each transaction record, $T_t$ represents the temporal information of user behavior, $T_a$ is the particular activity of the user at $T_t$ and $T_o$ is the other information related to $T_a$.

\textbf{Definition (Mobile phone data).} Mobile phone data $Db$ represents the behavioral transactions that are produced based on user's different activities with mobile phone over time. Let $BTs = \{BT_1,BT_2, ..., BT_m\}$ be a list of behavioral transactions related to a mobile phone user $U$, then $Db$ is a collection of $BT_i$ of size $m$.

\textbf{Definition (Base Period).} A base period $BP$ is a particular time duration that is used to capture the base behavioral pattern of the user.

\textbf{Definition (Time Slice).} A time slice $TS$ is a time boundary of a base period $BP$. Let $t_1$ is the start time point of $TS$ and $t_2$ is the end time point of $TS$, then $TS = [t_1, t_2]$, where $|(t_1 - t_2)| = BP$.

\textbf{Definition (Dominant Behavior).} The dominant behavior $D$ of a user $U$ in a particular time slice $TS$ is a particular activity that most commonly occurs among a list of activities in that time slice $TS$ by taking into account the data instances of different weeks considering the whole time period being a week. Let  $Oc = \{Oc_1,Oc_2, ..., Oc_n\}$ be a list of behavioral occurrences in percentage (\%) and n is the number of behavior classes in $TS$, then $D = MAX({Oc_1,Oc_2,...,Oc_n})$ is the dominant behavior of that $TS$.

\textbf{Definition (Time-Series-Segmentation).} A time-series segmentation is a process of transforming an input time-series continuous attribute $A$ into a sequence of $k$ discrete segments $ <{Seg_1, Seg_2, ..., Seg_k}>$ of disjoint intervals ${[t_0+1, t_1], [t_1+1, t_2],..., [t_{k-1}+1, t_k]}$, where $t_0$ is the minimal value, $t_k$ is the maximal value and $t_{i-1} < t_{i}$, for i = 1,2,..,k. 

Such intervals are produced in a way that the similar behavioral time-series are grouped together sequentially such that $[24$-$hours$-$a$-$day] = \cup^{k}_{i=1}Seg_i $ based on a certain similarity measure. The intervals $<{[t_0+1, t_1], [t_1+1, t_2],..., [t_{k-1}+1, t_k]}>$ are called segments, the times $<t_0, t_1,..., t_k>$ are called segment boundaries and $k$ indicates the number of segments.
 
\textbf{Definition (Temporal Behavior Rule).} A temporal behavior rule is an implication $X \rightarrow Y$, where $X$ contains temporal information $\{X \in \cup^{k}_{i=1}Seg_i$ and $[24$-$hours$-$a$-$day] = \cup^{k}_{i=1}Seg_i$\} of the week and $Y$ is the corresponding behavior of the user. The former, $X$, is called the antecedent of the rule, and the latter, $Y$, is called the consequent. Such temporal rules can be used to model individual's daily behavior for different purposes based on time-series data. \\

\textbf {Problem Formulation.} With the above definitions, the main problem we are addressing in this paper is formulated as follows: 

Given a user's mobile phone log dataset $Db$, our goal is to extract $k$ similar behavioral time segments from time-series data in $Db$ so that $\{[24$-$hours$-$a$-$day] = \cup^{k}_{i=1}Seg_i\}$  by calculating the number of optimal segments dynamically for each user $U$ without any prior knowledge and finally express these segments as temporal rules ($X \rightarrow Y$) in order to mine mobile phone user behavior. In this paper, we introduce a behavior-oriented segmentation technique for solving this problem.

\section{Our Approach}
\label{Our Approach}
In this section, we present our behavior-oriented time segmentation approach step-by-step for extracting temporal behavior rules, in order to mine individuals' behavior utilizing their mobile phone data. 

\subsection{Approach Overview}
\begin{figure}[H]
	\centering
	\includegraphics[width=\linewidth, keepaspectratio]{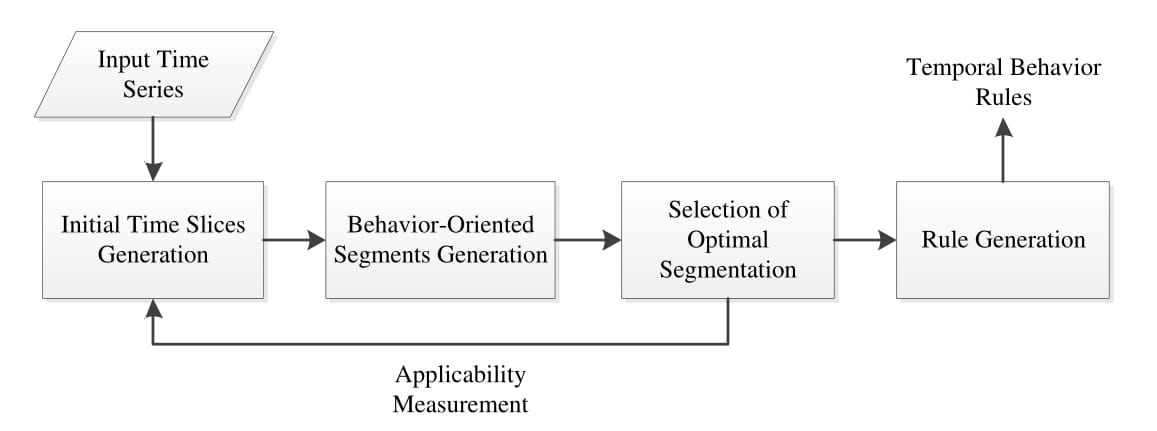}
	\caption{Approach overview}
	\label{fig:bots-overview}
\end{figure}

First, we generate initial time slices. For this, we divide each day of the week into relatively small time slices using a small \textit{base period}. For the purposes of this study, we assume a 5 minute period as the finest granularity required to distinguish day-to-day activities of an individual user. Second, we generate behavior-oriented segments. For this, we identify the dominant behavior of each slice and aggregate adjacent slices dynamically with the same dominant characteristics to get larger segments of similar behavior. These aggregated segments will have more support and temporal coverage and can be used as the basis for mining rules pertinent to the individuals. Third, we select optimal segmentation. For this, we measure the \textit{applicability} for that segmentation. As we have no prior knowledge about individual's behavioral patterns, we then iteratively increment the base period $(BP \times iteration \textit{++})$ and compare the applicability of the corresponding segmentation over each iteration in order to identify optimal base period. The time segmentation that yields the maximum applicability establishes the \textit{optimal time segmentation} and the corresponding base period is the optimal base period that captures the unique behavioral patterns of individuals. Finally, we generate the temporal rules using the discovered optimal segmentation for the users. Figure \ref{fig:bots-overview} shows the block diagram of the proposed segmentation approach for extracting temporal behavior rules of individuals. 

As individuals' behaviors differ from day-to-day (Section \ref{Introduction}), we take into account day-wise segmentation to better capture their daily behaviors. To achieve our goal, we initially split the whole log data into day-wise data and apply the segmentation technique on each set of day-wise data. Finally, the produced temporal rules are merged to get a complete set of rules that reflect day-wise behaviors (on a weekly basis) for individual users. In the following subsections, we describe the components of the above diagram one by one.

\subsection{Initial Time Slices Generation}
As our approach is individual's behavior-oriented, the first phase of our approach is initial time slices generation during the whole 24-hours-a-day time period for capturing the behavior of an individual. To do this, we initially divide each day of the week into relatively small time slices according to a base period. These initial time slices are used to capture the behavioral patterns of individuals because their daily behavior occurs in a time interval rather than at an exact time. The number of time slices depends on the length of the base period. If $T_{max}$ represents the whole time period of 24-hours-a-day and $BP$ is a base period, then the number of slices is - 

\begin{equation}
\label{Number of Slices}
	\textit{Number-of-Slices} = \frac{T_{max}}{BP}
\end{equation}

\begin{figure}[H]
	\centering
	\includegraphics[width=\linewidth, keepaspectratio]{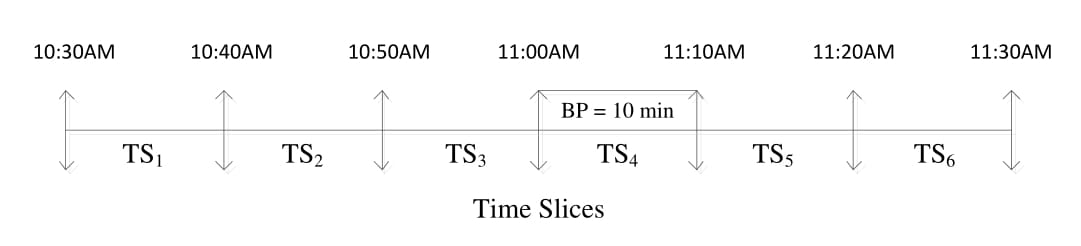}
	\caption{Initial time slices}
	\label{fig:time-slices}
\end{figure}

According to the equation (\ref{Number of Slices}), if the base period increases, the number of time slices decreases. For example, if the initial base period is 5 minutes, then the number of slices is $(24$-$hours$-$a$-$day) / 5 = 288$. A base time period, e.g., 5 minutes, is assumed as the finest granularity to distinguish day-to-day activities of an individual. If the base period incremented to $(5 \times 2)= 10$ minutes in second iteration, then the number of slices will be $(24$-$hours$-$a$-$day) / 10 = 144$. Figure \ref{fig:time-slices} shows an example of initial time slices $(TS_1,...,TS_6)$ including time boundaries of each slice between 10:30AM and 11:30AM when the base period $(BP)$ is 10 minutes.

\subsection{Behavior-Oriented Segments Generation}
\subsubsection{Dominant Behavior Identification}
In this step, we first identify the dominant behavior for each time slice generated in earlier phase as we take into account the \textit{diverse behaviors} of individuals over time. Dominant behavior represents the ``maximum number of occurrences'' of a particular activity among a list of activities in a time slice by taking into account the data instances of different weeks. As the pattern of an individual's behavior varies according to the duration of the regular activities they undertake during the week, we group the activity instances from the log into time slices. In this regard, we consider the whole time period being a week, i.e., assuming individual's regular behaviors follow a weekly pattern. As such, activities from different weeks for the same weekly time slices are merged, and the whole week is divided into consecutive time slices.

Therefore, the time slice that contains the dominant behavior can play a role to produce high confidence rule with that dominant behavior. As we have no prior knowledge about individual's behaviors over time-of-the-week, we may not get dominant behavior in some time slices. 

Assume that we have a time slice $TS_{30}$, with the following behavioral information, where the first parameter represents user behavior class and second parameter denotes the corresponding occurrences (\%) in $TS_{30}$. \\

\{  $TS_{30}: {(BH_1, 45\%),(BH_2, 45\%),(BH_3, 10\%)}$ \} \\

However, there is no dominant behavior in $TS_{30}$ as both $BH_1$ and $BH_2$ have the same number of occurrences (45\%). Therefore if we take into account $TS_{30}$ for producing rules, we get multiple rules with conflict behaviors ($BH_1$ and $BH_2$) that is impractical. In terms of rule's confidence, we can avoid such type of conflicting rules by taking into account more than 50\% occurrences for a particular behavior in a time slice.

Assume that we have another time slice $TS_{35}$, with the following behavioral information, where the parameters represent user behavior class and corresponding occurrences (\%) respectively in $TS_{35}$. \\

\{  $TS_{35}: {(BH_1, 55\%),(BH_2, 40\%),(BH_3, 5\%)}$ \} \\

Hence, $BH_1$ is the dominant behavior in $TS_{35}$ as $BH_1$ has the highest occurrences (55\%) comparing to others. As such, the time slice $TS_{35}$ can play a role to produce a conflict-free rule with the dominant behavior $BH_1$ that is meaningful. However, as we mentioned in (Section \ref{Introduction}), the confidence threshold for creating rules will vary according to an \textit{individual's preference} as to how interventionist they want the agent to be. Lets consider, the preferred confidence threshold is 75\% for a particular user U, e.g., he is not interested with those rules that have confidence less than 75\%. In that case, the produced rule using the time slice $TS_{35}$ will be meaningless for U, even though there is a clear dominant behavior ($BH_1$) in that time slice.

\begin{figure}
	\centering
	\includegraphics[width=\linewidth, keepaspectratio]{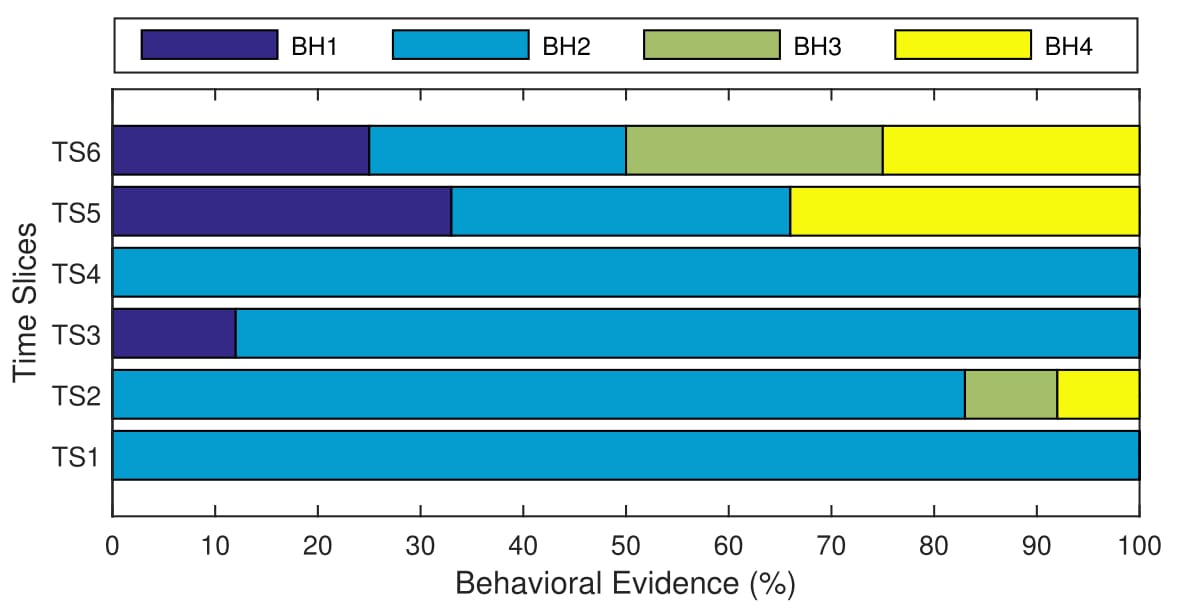}
	\caption{Sample behavioral data (\%) in different time slices}
	\label{fig:dominant-identification}
\end{figure}

Therefore, in order to produce behavioral rules according to the preferences of individuals, we use the preferred rule confidence threshold (t) to identify the dominant behavior of each time slice. The benefit of using this threshold is that it reduces the burden of processing to get the expected segmentation according to individuals' preferences. In a time slice, if the percentage of a particular behavior class $BH_i \geq threshold (t)$ then $BH_i$ is the dominant behavior for that time slice. Figure \ref{fig:dominant-identification} shows a sample behavioral data evidence for identifying dominant behavior for different time slices assuming the preferred confidence threshold 75\%. 

According to Figure \ref{fig:dominant-identification}, $TS_1$ contains 100\% $BH_2$ that satisfies the threshold, so $BH_2$ is the dominant behavior for this slice. $TS_2$ contains 83\% $BH_2$, 8\% $BH_3$, and 9\% $BH_4$, so $BH_2$ is the dominant behavior for this slice as it also satisfies the threshold. Similarly, $BH_2$ is the dominant behaviors for time slices $TS_3$ and $TS_4$ as well. However, there is no dominant behavior for the time slices $TS_5$ and $TS_6$ because of not getting any behavior greater than 75\%. As the dominant behavior represents the highest number of occurrences of a particular behavior, maximum one dominant behavior is identified in a slice. If $TS_{total}$ represents the total number of time slices then the number of time slices that contain the dominant behavior is -
\begin{equation}
	\textit{Number-of-TS(dominant)} \leq TS_{total}
\end{equation}

\subsubsection{Dynamic Aggregation}
In our technique, once the dominant behavior has been identified for each time slice, slices that exhibit same dominant behavior are dynamically aggregated into longest possible time segments. This is done to increase the \textit{support value} and \textit{temporal coverage} for any rules that are eventually extracted for these time segments. 

Assume that we have four consecutive time slices $TS_{1}, TS_{2}, TS_{3}$ and $TS_{4}$, with the following behavioral information (Shown in Figure \ref{fig:dominant-identification}), where the first parameter represents the time slice and second parameter denotes the corresponding dominant behavior for that time slice. \\

\{${(TS_{1}, BH_2),(TS_{2}, BH_2),(TS_{3}, BH_2), (TS_{4}, BH_2)}$\} \\

\begin{figure}
	\centering
	\includegraphics[width=\linewidth, keepaspectratio]{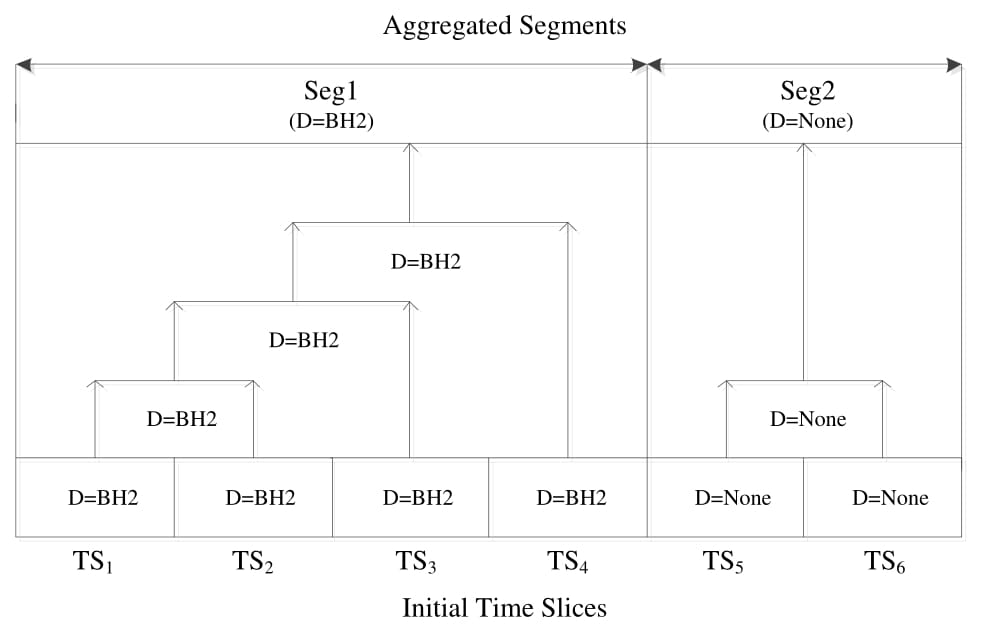}
	\caption{Dominant behavior based dynamic aggregation of initial time slices}
	\label{fig:dynamic-aggregation}
\end{figure}

As each of these time slices has the dominant behavior, these slices are able to produce meaningful rules separately in terms of confidence. However, in order to get an \textit{effective behavior-oriented segment}, we aggregate these time slices into one single longest segment $Seg_1$ (Shown in Figure \ref{fig:dynamic-aggregation}) as they contain same dominant behavior. As such, this longest similar behavioral segment is able to produce more meaningful rule in terms of support, temporal coverage and confidence with the dominant behavior $BH_2$. 

In order to discover such longest similar behavioral segments, we use bottom-up hierarchical aggregation technique based on dominant behavior. The most similar technique is agglomerative clustering algorithm \cite{xu2005survey} that use a proximity matrix which is generated by computing the distance between clusters. According to the matrix value the algorithm successively merge the clusters until the desired cluster structure is obtained that is defined by a threshold. However, it is very difficult to predict the threshold level at which the merging is best according to a proximity matrix because of the variations in users' behavior over time. Therefore, we produce consecutive segments by aggregating initial time slices dynamically based on dominant behavior, in which some segments are produced using more merging and other segments are produced using less merging, depending on the changes in individual mobile users' behavior. Figure \ref{fig:dynamic-aggregation} shows a sample example of producing such dynamic segments [${Seg_1, Seg_2}$] from the initial time slices using dynamic aggregation where $BH2$ is the dominant behavior of ${Seg_1}$ $[D=BH2]$ and ${Seg_2}$ $[D=None]$ has no dominant behavior.

\begin{algorithm}[h]
	\caption{Dynamic Aggregation}
	\label{alg:Dynamic-Aggregation}
	\DontPrintSemicolon
	\SetAlgoLined
	\SetKwInOut{Data}{Data}
	\nl \Data{initial time slices list: $TS_{list}$}
	\nl \KwResult{behavior-oriented segment list: $Seg_{list}$}
	
	\BlankLine
	
	//create initial segment using the first time slice\\
	\nl $Seg_{init} \leftarrow TS_{1}$ \\
	//insert segment into the segment list\\
	\nl $Seg_{list} \leftarrow insert (Seg_{init}$)\\
	\nl \ForEach{$TS$ in $TS_{list}$}
	{
		//identify dominant behavior using the threshold t\\
		\nl $D \leftarrow identifyDominant (TS, t)$ \\
			//check the dominant behavior \\
		\nl \eIf{ $D(Seg_{init}) \equiv D(TS)$}
		{
				//aggregate into one segment \\
			\nl $Seg_{agg} \leftarrow$ aggregate ($Seg_{init}, TS$) \\
				//initial segment is changed to aggregated segment\\
			\nl $Seg_{init} \leftarrow Seg_{agg}$ \\
				//update segment list\\
			\nl $Seg_{list} \leftarrow update (Seg_{init}$)
		}
		{
				//create new segment using the next time slice\\
			\nl $Seg_{new} \leftarrow$ createSeg($TS$) \\
				//insert segment into the list \\
			\nl $Seg_{list} \leftarrow insert (Seg_{new}$)
		}
		
	}
	\nl return $Seg_{list}$
\end{algorithm}

The process for doing this dynamic aggregation is set out in Algorithm \ref{alg:Dynamic-Aggregation}. Input data includes initial time slices list $TS_{list}$ (line 1) and output data is the list of behavior-oriented segments $Seg_{list}$ (line 2). A segment $Seg_{init}$ is initialized using the first time slice $TS_{1}$ (line 3). For each time slice, the method $identifyDominant()$ identifies the dominant behavior using the threshold t (line 6). After that we check the dominant behavior of $TS$ and $Seg_{init}$ (line 7). If the same dominant found then we aggregate these two time slices into one segment by updating the contents and time boundaries (line 8). After that initial segment is changed to aggregated segment and we update the segment list as well. This aggregation continues until different dominant behavior found is encountered in $TS_{list}$. When the different dominant is found we then create a new segment $Seg_{new}$ (line 11) and insert into the segment list (line 12) and continue aggregating with this new segment by similar manner. In this way, some segments are produced by aggregating large number of $TS$ (e.g., segment $Seg_1$ in Figure \ref{fig:dynamic-aggregation}) while some may have a smaller number of $TS$ (e.g., segment $Seg_2$ in Figure \ref{fig:dynamic-aggregation}) depending on how the user's behavior changes over time.

Rather than arbitrarily determine the number of segments in advance, our algorithm dynamically derives the number of segments to be produced from an individual's data. Thus the number of segments and time boundaries of the produced segments will differ from user-to-user.

\subsection{Selection of Optimal Segmentation}
\subsubsection{Segments Filtering}
As different length of segments with different dominant behaviors (For example, $Seg_1$ with $D$=$BH2$ and $Seg_2$ with [$D$=$None$], shown in Figure  \ref{fig:dynamic-aggregation}) are produced after performing dynamic aggregation, we need to select segments that are able to produce high confidence temporal rules to reduce the burden of the processing. The reason is that it is unlikely to get behavioral rules using all the segments generated by dynamic aggregation as individual's behavior is not consistent over time-of-the-week in the real world. 

To select segments that are able to produce behavioral rules according to the preferred confidence of individuals, we simply ignore those segments that have no particular dominant behavior, (e.g., segments with [$D=None$]). Because there is no possibility to produce temporal rules that satisfy the user preferred confidence using the segments having [$D=None$]. Therefore, we keep only the segments that have a particular dominant behavior in order to produce meaningful temporal behavior rules of individuals.

Assume that we have three segments with the following behavioral information, where the first parameter represents time segments and second parameter denotes the corresponding dominant behavior after dynamic aggregation. \\

\{$(Seg_{1}, BH_2), (Seg_{2}, None), (Seg_{3}, BH4)$ \} \\

As $Seg_{2}$ $[D=None]$ has no dominant behavior, this segment is unable to produce any meaningful behavioral rule according to the individual's preference. Therefore, we reduce the segments size by filtering such segments and take into account $Seg_{1}$ and $Seg_{3}$ for producing rules, as each of these segments contain particular dominant behavior that is the basis for producing effective behavioral rules of the users. 

\subsubsection{Applicability Measurement}
Different base periods may give different time segmentation and related rules, due to their impact on support, temporal coverage and confidence. As all the filtered segments having the dominant behavior are able to produce rules according to individual's preference, we assume each of such segments as an antecedent of the temporal rule for measuring applicability. 

In order to identify the \textit{optimal segmentation}, we propose a metric `applicability' that measures the \textit{applicability} of rules generated by the above filtered segments having a particular dominant behavior. \textit{Applicability} is a descriptive statistic that takes into account two parameters for a particular confidence threshold. These are:

\begin{enumerate}[(i)]
  \item Temporal coverage - is the time interval covered by a temporal rule. If $t_{start}$ and $t_{end}$ is the start and end time point of a particular time segment that is used to produce a temporal rule R, then the temporal coverage for that rule $Rcov = |t_{end} - t_{start}|$, e.g., the internal time interval of that segment.
  \item Support - is the number of behavioral instances ($Rsup$) in a time segment that is used to produce a temporal rule.
\end{enumerate}

In our approach, segmentation of time over the week is taken as the proxy of the user's activities and subsequent behavior. On one hand we want time segmented with enough resolution to discriminate between various types of dominant behavior for a particular confidence threshold. We also want rules that capture that behavior to have as much support as possible. However, the metric `confidence' and `support' of association rule learning \cite{agrawal1994fast} are not sufficient for identifying optimal temporal rules in order to mine mobile user behavior. The reason is - temporal rules may have the \textit{temporal coverage} either small or large that depends on the volatility of a user's behavior stability over time. The traditional metric takes into account each context (e.g., time segment having time interval small or large) as a particular item that is more meaningful in market basket analysis. Thus it does not reflect the effects of \textit{temporal coverage} in discovering meaningful behavioral rules of users.

\smallskip
We define our new `applicability' \textit{metric} as follows: 
\smallskip

\textit{Applicability}: It is defined as the product of aggregate support and aggregate temporal coverage, where \textit{aggregate support} is the fraction of the summation of the support count of all the rules that satisfies the confidence threshold among the maximum possible support considered and \textit{the aggregate temporal coverage} is the proportion of the temporal coverage by those rules. Formally, the applicability is defined as:

\begin{equation}
\label{applicability}
Applicability =\sum_{i=1}^N \left( \frac{Rsup_i}{S_{max}} * \frac{Rcov_i}{C_{max}} \right) 
\end{equation}

where, $Rsup$ is the support count of a rule, $Rcov$ is the temporal coverage of the rule, $S_{max}$ is the maximum possible support in a dataset, $C_{max}$ is the maximum possible temporal coverage in a week and `N' is the number of rules that satisfies the user's confidence threshold.

\subsubsection{Identify Optimal Segmentation}
As discussed above, the applicability of temporal rules for a particular confidence threshold is dependent on the produced dynamic segments list that is based on the length of \textit{base period}. The most appropriate segmentation will depend on the particular pattern of the user's diverse behaviors. As we have no prior knowledge about individual's behavioral patterns, we then iteratively increase the base period by a reasonable time gap and compare the applicability of the corresponding segmentation over each iteration in order to identify optimal base period. The time segmentation that yields the maximum applicability establishes the \textit{optimal time segmentation} and the corresponding base period is the \textit{optimal base period} that captures the unique behavioral patterns of individuals. As our approach is \textit{individualized behavior-oriented}, the optimal base period to capture the behavioral pattern and corresponding optimal segments for producing temporal behavior rules vary from user-to-user.

\begin{algorithm}[h]
\resetlinenumber
	\caption{Identify Optimal Segmentation}
	\label{alg:optimal-segmentation}
	\SetKwInOut{Data}{Data}
	\SetAlgoLined
	\nl \Data{base period: $BP$}
	\nl \KwResult{optimal segments list: $OSeg_{list}$}
		
	\BlankLine
		
	//initialize applicability \\
	\nl $A_{init}$ $\leftarrow$ 0 \\
	\nl \ForEach{$BP$ in 24-hours-a-day time scale}
	{  
		 //generate initial time slices using base period\\
		\nl   $TS_{list} \leftarrow$ generateTS(BP) \\
		//produce behavior-oriented aggregated segments\\
		\nl   $Seg_{list} \leftarrow$ aggregateSeg($TS_{list}$) \\
		//get filtered segments\\
		\nl   $FSeg_{list} \leftarrow$ filterSeg($Seg_{list}$) \\
		//calculate the applicability utilizing filtered segments\\ 
		\nl $Applicability$ $\leftarrow$ calculateApplicability($FSeg_{list}$) \\
		//compare the applicability\\
		\nl \If{$Applicability$ $>$ $A_{init}$}
		{
		//store the base period as optimal base period\\
		 \nl $BP_{optimal} \leftarrow BP$ \\
		//update initial applicability\\
		 \nl $A_{init} \leftarrow Applicability$ \\
		//update optimal list\\
		\nl $OSeg_{list} \leftarrow$ updateOSegList($Seg_{list}$)
		}
		//next base period\\
	\nl	increase $BP$
	}	
	\nl	return $OSeg_{list}$
\end{algorithm}

The overall process is shown in Algorithm 2. Input data includes base period BP (line 1) and output data is the list of optimal segments $OSeg_{list}$ (line 2). Applicability $A_{init}$ is initialized to zero (line 3). For each base period, the method $generateTS()$ generates initial time slices $TS_{list}$ using the base period $BP$ (line 5), after that the method $aggregateSeg()$ produces behavior-oriented segments $Seg_{list}$ by aggregating similar behavioral segments (line 6). As all the aggregated segments are not able to produce high confidence temporal rules, we select segments that contains particular dominant behavior using the method $filterSeg()$ (line 7). We then calculate the applicability $Applicability$ using the filtered segments in method $calculateApplicability()$ (line 8). The applicability is then compared with the initial applicability $A_{init}$ (line 9). If greater applicability is found then we store the base period $BP$ as an optimal base period $BP_{optimal}$ (line 10), after that initial applicability $A_{init}$ is changed to new applicability $Applicability$ for the purpose of comparing in the next iteration (line 11) and update optimal segments list $OSeg_{list}$ with $Seg_{list}$ (line 12). By increasing base period $BP$, we continue this process (line 13) to identify the optimal base period and corresponding segments. Finally, this algorithm returns the optimal segments list $OSeg_{list}$ that is generated for the optimal base period $BP_{optimal}$ (line 14). 

\subsection{Rule Generation}
In order to produce the temporal rules of an individual user utilizing the optimal segmentation we employ the well-known association rule learning algorithm Apriori \cite{agrawal1994fast}. A key benefit of using association rule learning is that a discovered behavioral rule will have a high predictive accuracy \cite{freitas2000understanding} as it allows an individual for creating rules according to her preference. Moreover, it can be easily read and understood by both the end user and the developer \cite{srinivasan2014mobileminer}. 

A temporal rule is represented as  $X \rightarrow Y$, where X is defined as the antecedent and Y as the consequent. The algorithm generates rules with the antecedent containing temporal information [day-of-the-week, time segment] and consequent containing only individual's behavior at that time period. This means that rules can be in the form $X \rightarrow Y$ but not in the form of $Y \rightarrow X$. To better understand the concept of temporal rules let us consider an example of phone call behaviors where the user: (i) always makes outgoing calls between 13:00 and 14:00 on Thursdays; (ii) rejects the incoming calls between 14:10 and 15:35 on Mondays; (iii) misses most of the incoming calls between 19:00 and 20:00 on Saturdays, then the following temporal rules would represent the user's preferences in this case: 

\begin{equation*}
	\begin{split}
	& (i) Thursday [13:00-14:00] \Rightarrow Outgoing \\
	& (ii) Monday [14:10-15:35] \Rightarrow Reject \\
	& (iii) Saturday [19:00-20:00] \Rightarrow Missed
	\end{split}
\end{equation*}

The algorithm scans the data and produce such temporal rules by checking the parameters `support' and `confidence' that is defined as:

\begin{itemize}
	\item \textit{Support}: the ratio between the number of times X and Y co-occur and the number of data-instances present in the given data. It can be represented as the joint probability of X and Y : $P(X,Y)$.
	
	\bigskip \item \textit{Confidence}: the ratio between the number of times Y co-occurs
	with X and the number of times X occurs in the given data. It can be represented as the conditional probability of X and Y : $P (Y | X)$.
\end{itemize}

A temporal rule is created only when it has at least the minimum support and confidence. It is worth noting that decreasing the values of either support or confidence could result in discovering more rules \cite{agrawal1994fast}.
		
\section{Experiments}
\label{Experiments}
To validate our BOTS approach, we have conducted a range of experiments on the real mobile phone datasets for mining temporal behavior rules of individual mobile phone users. We have implemented both our BOTS approach and existing approaches in Java programming language and executed them on a Windows PC with an Intel Core I5 CPU (3.20GHz) and 8GB memory. In the following subsections, we briefly describe the datasets, and present the experimental results and discussion.

\subsection{Datasets}
In our experiments, we have used two different datasets that include the temporal information and corresponding behavior of individuals. These are:

\subsubsection{Reality-Mining Dataset}
This dataset consists of 94 individual mobile phone users over nine months which were collected at Massachusetts Institute of Technology (MIT) by the Reality Mining Project [Massachusetts Institute of Technology 2007] \cite{eagle2006reality}. These 94 individuals are faculty, staff, and students. The datasets include people with different types of calling patterns and call distributions. We extract 5-tuple information of the call record related to temporal information and corresponding behavior for each phone user from the datasets: {Date of call, Time of call, Type of call, Call duration, Call ID}. This dataset contains three types of phone call behavior, e.g., INCOMING, MISSED and OUTGOING. As can be seen, the user's behavior in ACCEPTing and REJECTing calls are not directly distinguishable in INCOMING calls in the dataset. As such, we derive ACCEPT and REJECT calls by using the call duration. If the call duration is greater than 0 then the call has been ACCEPTED; if it is equal to 0 then the call has been REJECTED \cite{sarker2016behavior}.

\subsubsection{Swin Dataset}
This dataset was collected directly from individual mobile phone users by us. To do this, we have first developed an Android mobile app which collects the user's real current call log data (Date of call, Time of call, User phone call behavior, Call ID) on their mobile phones. Using our app, data was collected from 22 individual mobile users of different professions such as undergraduate students, post graduate students, university lecturers and industry professionals, from August 2014 to September 2015. This dataset contains four different types of phone call behavior, e.g., ACCEPT, REJECT, MISSED and OUTGOING.

\subsection{Evaluation Metric}
In order to assess our behavior-oriented segmentation approach for extracting temporal behavior rules, we take into account the following measurements:

\begin{itemize}
	\item \textit{Applicability:} It measures not only the support of temporal rules but also the temporal coverage of those rules. According to equation (\ref{applicability}), it is the product of aggregate support and aggregate temporal coverage, where \textit{aggregate support} is the fraction of the summation of the support count of all the rules that satisfies the confidence threshold among the maximum possible support considered and \textit{the aggregate temporal coverage} is the proportion of the temporal coverage by those rules.

	\item \textit{Data Coverage and Accuracy:} Coverage measures the percentage of tuples that is covered by the produced segments and accuracy measures the percentage of tuples that is identified with correct behavior in a dataset. Given a class labeled dataset, $Db$, let $n_{covers}$ be the number of tuples covered by the segmentation;  $n_{correct}$ be the number of tuples correctly classified by the behaviors of that segmentation; and $|Db|$ be the number of tuples in $Db$. According to \cite{han2011data}, we can define the coverage and accuracy as -

	\begin{equation}
		Coverage = \frac{n_{covers}}{|Db|} * 100\%
	\end{equation}	
	\begin{equation}
		Accuracy = \frac{n_{correct}}{n_{covers}} * 100\%
	\end{equation}	
\end{itemize}

As the behavior-oriented segments are used to produce temporal rules, to assess individual's temporal behavior rules corresponding to that segmentation, we compare the predicted behavior with the actual behavior (i.e., the ground truth) and compute the accuracy in terms of:

\begin{itemize}
	\item Precision: ratio between the number of activities that are correctly predicted and the total number of activities that are predicted (both correctly and incorrectly). If TP and FP denote true positives and false positives then the formal definition of precision is:
		
		\begin{equation}
		Precision = \frac{TP}{TP + FP}
		\end{equation}
	
	\item Recall: ratio between the number of activities that are correctly predicted and the total number of activities that are relevant. If TP and FN denote true positives and false negatives then the formal definition of recall is:
		
		\begin{equation}
		Recall = \frac{TP}{TP + FN}
		\end{equation}
\end{itemize}

\subsection{Experimental Results and Discussion}
We report the overall results of our experiments on real mobile phone datasets and illustrate our approach with the detailed of experimental results of two individuals (randomly selected) from the above mentioned datasets. User 10 is selected from `Swin' dataset and User 51 is selected from `Reality-Mining' dataset.

\subsubsection{Individualized Time Segments and Corresponding Temporal Rules}  
In this experiment, we show individualized behavior-oriented segments and corresponding temporal behavior rules produced by our approach. For this, we initially split the whole log data into day-wise data and apply the segmentation technique on each set of day-wise data. Finally, we merge the produced temporal rules for individual users. Table \ref{Sample-mined-rules-call} shows sample phone call behavioral rules of individuals. As our approach produces behavioral rules for a particular preferred confidence threshold of individuals, the results are presented for a given confidence threshold 75\% (default setting). 

\begin{table}
	\centering
	\caption{Sample behavior-oriented segments and corresponding temporal behavior rules}
	\label{Sample-mined-rules-call}
	
	\begin{tabular}{|c|p{5cm}|c|} \hline
		\bf Users & \bf \centering Behavioral Rules & \bf Confidence \\ \hline 
		 & $Day \rightarrow Saturday, TimeSegment \rightarrow [19:00-20:00]
		\Rightarrow Behavior \rightarrow Missed$ & 85\% \\  \cline{2-3}
		User 10 & $Day \rightarrow Thursday, TimeSegment \rightarrow [13:00-14:00]
		\Rightarrow Behavior \rightarrow Outgoing$ & 100\% \\ \hline 
		
		& $Day \rightarrow Friday, TimeSegment \rightarrow [21:30-22:30]
		\Rightarrow Behavior \rightarrow Accept$ & 88\% \\ \cline{2-3}
		User 51 & $Day \rightarrow Monday, TimeSegment \rightarrow [14:10-15:35]
		\Rightarrow Behavior \rightarrow Reject$ & 75\% \\  \hline
	\end{tabular}
\end{table}

If we observe Table \ref{Sample-mined-rules-call}, we see that User 10 misses most of the calls (85\%) between 19:00 and 20:00 on Saturdays and always (100\%) makes outgoing calls between 13:00 and 14:00 on Thursdays. On the other hand, User 51 accepts most of the calls (88\%) between 21:30 and 22:30 on Fridays and rejects most of the calls (75\%) between 14:10 and 15:35 on Mondays. The results in Table \ref{Sample-mined-rules-call} show that different users do have different behavior-oriented time segments and corresponding individualized rules.

\subsubsection{Effect of Base Period}
\begin{figure}
	\centering
	\includegraphics[width=\linewidth, keepaspectratio]{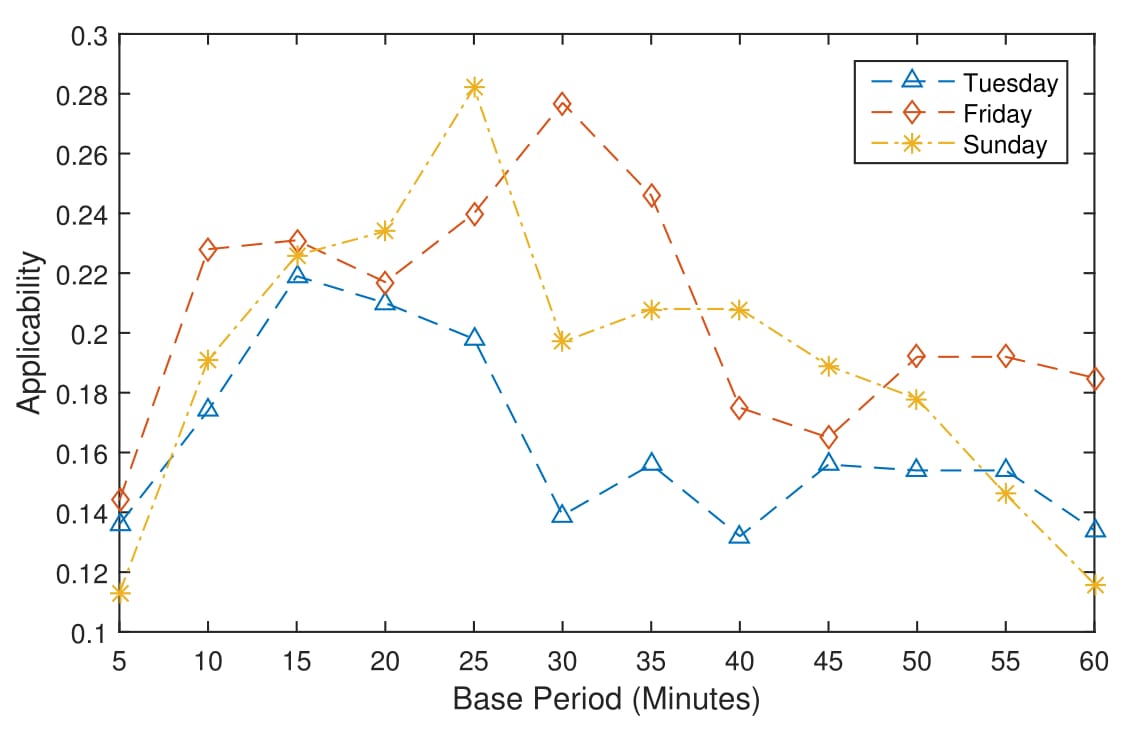}
	\caption{Effect of different base periods on segmentation quality (Optimal base period selection for different days-of-the-week of a sample user [User 51])}
	\label{fig:base-period-effect}
\end{figure}

In this experiment, we show the effect of base period on segmentation and on individuals as well. To show the effect of base period on segmentation, first we illustrate the detailed outcomes by varying the base periods for an individual user. In our experiment, initially we consider 5 minutes (reasonable small duration) as base period and then we iteratively increase by 5 minutes as a reasonable time gap to capture the behavior pattern of the user. The corresponding applicability for these base periods are compared. Figure \ref{fig:base-period-effect} presents the impacts of base periods on applicability (up to 60 minutes) for different days (randomly selected) Tuesday, Friday and Sunday respectively, for a particular confidence threshold 75\%. The x-axis of the figure is the base periods (in minutes) and y-axis represents the corresponding applicability for the behavior patters of different days. 

If we observe Figure \ref{fig:base-period-effect}, we can see that initially the applicability is low, it increases up to a certain base period, and then it again decreases. The reason is that if the initial time slices are small periods, the aggregate support and aggregate temporal coverage of produced rules will be very small and the resulting applicability is consequently small. On the other hand, if the initial time slices are large periods, some diverse behaviors within a slice will mask the dominant behavior and lose overall significance by producing rules with low confidence, resulting in such rules not being considered because of not satisfying the confidence threshold. As a result, the overall applicability is reduced.

The base period that produces the highest (peak) applicability for a particular confidence threshold, is the \textit{optimal base period}. From Figure \ref{fig:base-period-effect}, we found that for Tuesday, 15 minutes is the optimal base period that produces the maximal (peak) applicability. In other words, the initial time slices using 15 minutes base period is the best to capture the behavior pattern of Tuesday for this user. Similarly for Friday and Sunday, the applicability is maximal (peak) when the base period is 30 minutes and 25 minutes respectively. If we observe Figure \ref{fig:base-period-effect}, we see that the optimal base period for capturing behavioral patterns of an individual is not identical for all days-of-the-week, it differs from day-to-day of the week. The reason is that the user has different behavior patterns in different days-of-the-week.

\begin{figure}
	\centering
	\includegraphics[width=\linewidth, keepaspectratio]{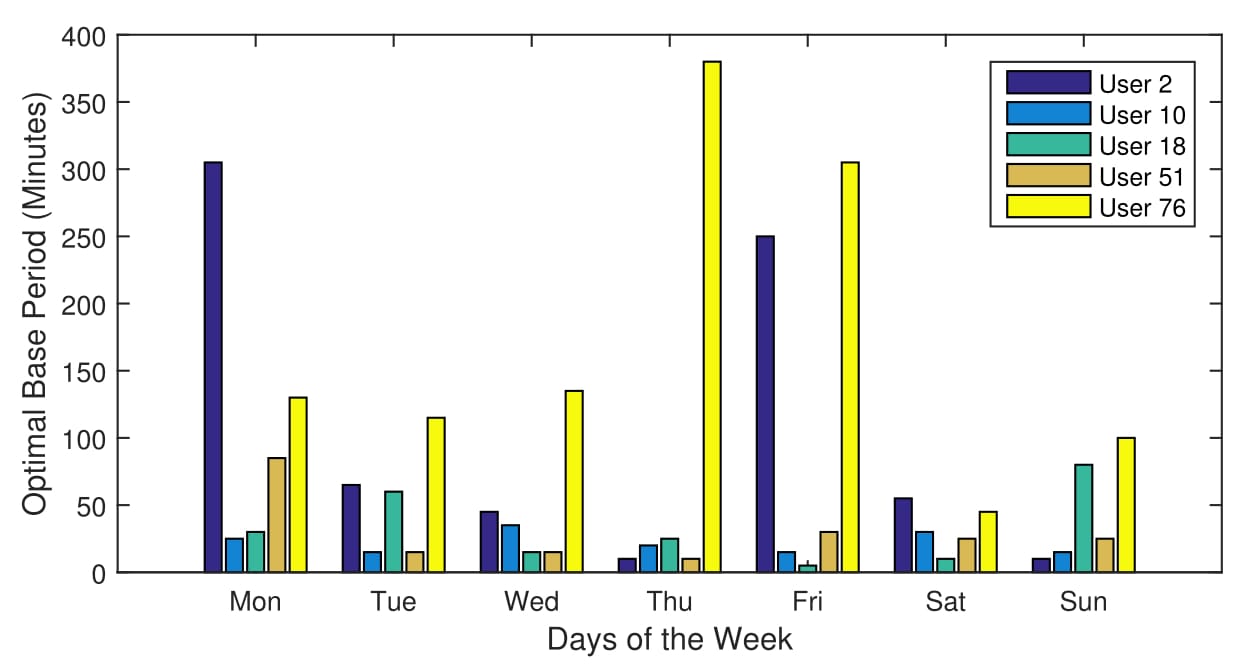}
	\caption{Effect of optimal base period on different individuals for different days-of-the-week}
	\label{fig:optimal-base-period-individuals}
\end{figure} 

As the behaviors of all individuals are not identical in the real word, these optimal base periods differ from user-to-user as well. To show the effect of optimal base period on individuals, Figure \ref{fig:optimal-base-period-individuals} reports the optimal base periods (OBP) discovered for five different individuals (randomly selected) by conducting experiments on their mobile phone data using same confidence threshold 75\%. If we observe Figure \ref{fig:optimal-base-period-individuals}, we see that the optimal base period for capturing behavioral patterns are not identical for all users, it differs from user-to-user. The reason is that different individuals have different behavior patterns in different days-of-the-week.

\subsubsection{Effect of Days-of-the-Week on Segmentation}
\begin{figure}
	\centering
	\includegraphics[width=\linewidth, keepaspectratio]{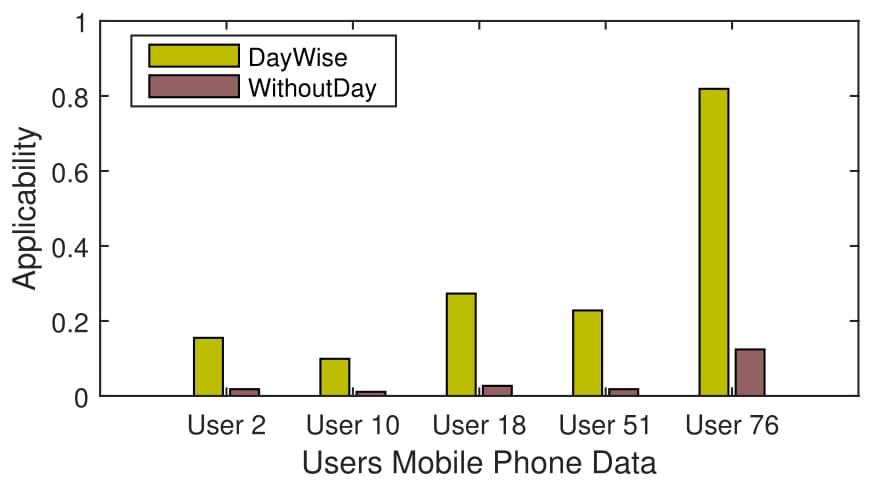}
	\caption{Effect of days-of-the-week on segmentation for different Individuals}
	\label{fig:day-effect}
\end{figure}

In this experiment, we show the effect of days-of-the-week on time segmentation. Figure \ref{fig:day-effect} shows the comparison of applicability by taking into account both day-wise segmentation and without-day-wise segmentation for different individuals.

If we observe Figure \ref{fig:day-effect}, we see that the applicability is higher when taking into account day-wise segmentation for different individuals. The reason is that, for many users their daily schedule differ from day-to-day. For instance, a user has a meeting on every Monday during [2:00PM-3:00PM] and rejects (not answer) the incoming calls during that time, but on other days he has no scheduled event at that time and accepts (answer) the incoming calls. Therefore, to capture such diverse behaviors in different days, day-wise behavioral patterns are needed to take into account. The results in Figure \ref{fig:day-effect} shows that day-wise segmentation is more meaningful to capture the daily behavioral patterns of individuals for mining behavioral rules.

\subsubsection{Effect of Execution Time on Data Size}
\begin{figure}
	\centering
	\includegraphics[width=\linewidth, keepaspectratio]{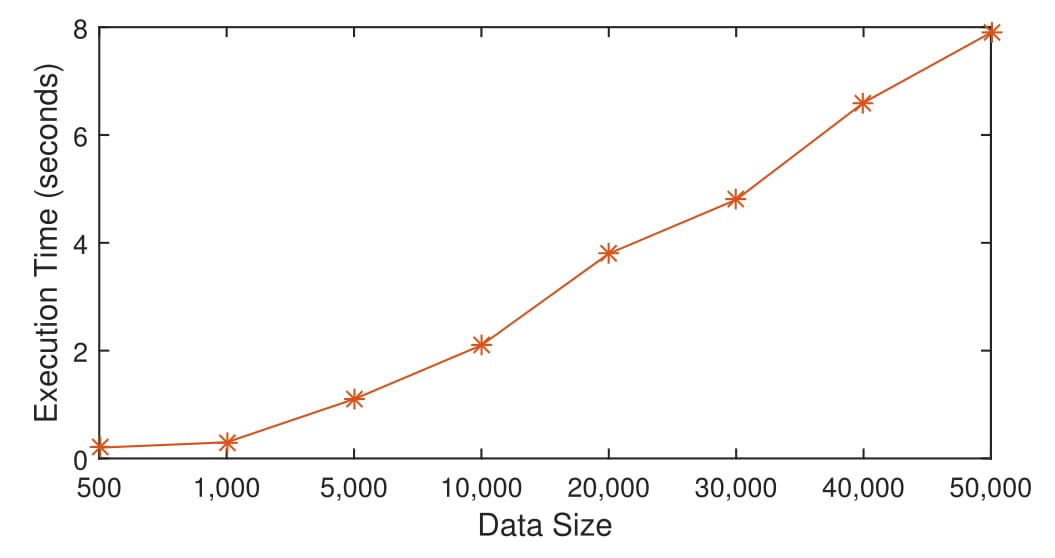}
	\caption{Effect of execution time on different data sizes}
	\label{fig:execution-time-effect}
\end{figure}

As we choose iterative process for identifying the optimal base period in our approach, to show the effect of execution time on data size, Figure \ref{fig:execution-time-effect} shows the execution time taken by our approach for different data sizes (from 500 instances to 50,000 instances).

If we observe Figure \ref{fig:execution-time-effect}, we see that our BOTS approach efficiently performs for different data sizes. To process, up to 5000 data instances, it takes only 1 second when executed them on a Windows PC with an Intel Core I5 CPU (3.20GHz) and 8GB memory. If the data size increases, it linearly increases the execution time. According to Figure \ref{fig:execution-time-effect}, to process 50,000 data instances of an individual user, our approach takes less than 8 seconds that ensures the efficiency of our approach.

\subsubsection{Effect of Confidence}
\begin{figure}
	\centering
	\includegraphics[width=\linewidth, keepaspectratio]{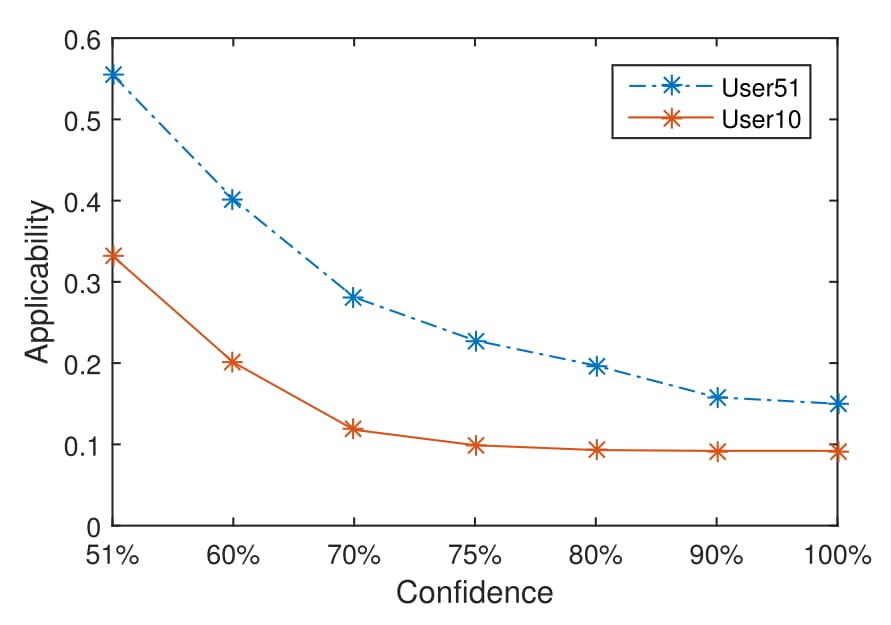}
	\caption{Effect of confidence on segmentation in terms of applicability for individual's mobile phone data}
	\label{fig:confidence-effect-on-applicability}
\end{figure}

\begin{figure}
	\centering
	\includegraphics[width=\linewidth, keepaspectratio]{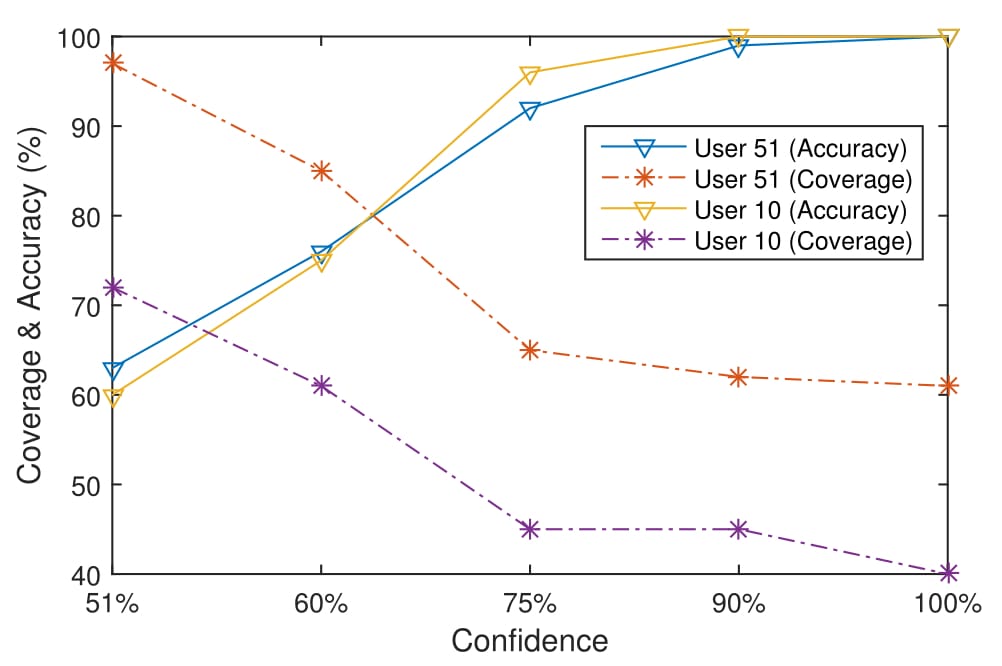}
	\caption{Effect of confidence on segmentation in terms of data coverage (\%) and accuracy (\%) on individual's mobile phone data}
	\label{fig:confidence-effect-on-coverage}
\end{figure}

In this experiment, we show the effect of confidence on segmentation and corresponding temporal rules. For this, we first illustrate the detailed outcomes by varying the conference threshold from 51\% (lowest) to 100\% (maximum) for different individuals. Since by the definition, confidence is associated to a \textit{rule's strength}, we are not interested to take into account below 51\% as confidence threshold. The reason is that below this confidence threshold, conflict behavior may be found for a particular temporal information that is impractical in rules. To show the effect of confidence on segmentation, Figure \ref{fig:confidence-effect-on-applicability} and Figure \ref{fig:confidence-effect-on-coverage} show the comparison of applicability, data coverage (\%) and accuracy (\%) for different confidence threshold for different individuals.

If we observe Figure \ref{fig:confidence-effect-on-applicability} and Figure \ref{fig:confidence-effect-on-coverage}, we see that applicability and coverage decreases with the increase of confidence threshold. The main reason for changing applicability with the confidence threshold is that our approach dynamically aggregates time segments with the dominance threshold being the same as the selected confidence threshold. Segments with the 51\% threshold are greater than those with 100\%, resulting in greater temporal coverage and greater support and therefore higher applicability. Similarly, data coverage (\%) also changes with the confidence threshold as coverage is directly associated with the percentage of data instances (support) covered by the produced segments in the dataset. On the other hand, accuracy increases with the increase of confidence threshold. If the confidence threshold is low, greater segments with greater behavioral variations are produced and the resulting accuracy is consequently low. On the other-hand, if the confidence threshold is high, comparatively smaller segments with less behavioral variations are produced and the resulting accuracy is consequently high, e.g., \textit{confidence represents the accuracy level}. The setting of this confidence threshold for creating rules will vary according to an individual's preference as to how interventionist they want the call-handling agent to be. The users need to choose a particular confidence threshold according to individual's preference (say 75\%), for generating their behavioral rules. 

As confidence is directly associated with accuracy, the applicability and data coverage (\%) ensure the quality of segmentation for mining rules for a particular confidence threshold (accuracy level). In the following subsection, we compare the applicability and data coverage (\%) for all techniques in order to show the effectiveness of our approach for different confidence threshold.

\subsubsection{Effectiveness Comparison}
In this experiment, we show the effectiveness of our BOTS approach in terms of applicability and data coverage (\%) comparing it existing time segmentation approaches. To do this, first we select 5 baseline methods that use different time segments for mining mobile user behavior. For comparison purposes, we denote these baseline methods as BM1 \cite{ozer2016predicting} that uses 15-minutes equal interval for time segmentation to mine human mobility patterns, BM2 \cite{mehrotra2016prefminer} that uses 4-unequal time slots based segmentation for learning mobile user preferences for notification management, BM3 \cite{zhu2014mining} that uses 5-unequal time slots for time segmentation for mining mobile user preferences for personalized recommendation, BM4 \cite{mukherji2014adding} that uses 4-hours equal interval based time segmentation for learning phone usages sequential patterns in order to build mobile sequence mining engine and finally BM5 \cite{phithakkitnukoon2010activity} that uses 3-hours equal interval for time segmentation to identify human daily activity patterns utilizing mobile phone data respectively. For these baseline techniques, we aggregated behaviors of different weeks utilizing the same datasets in order to compare the techniques fairly.

\begin{figure}
	\centering
	\includegraphics[width=\linewidth, keepaspectratio]{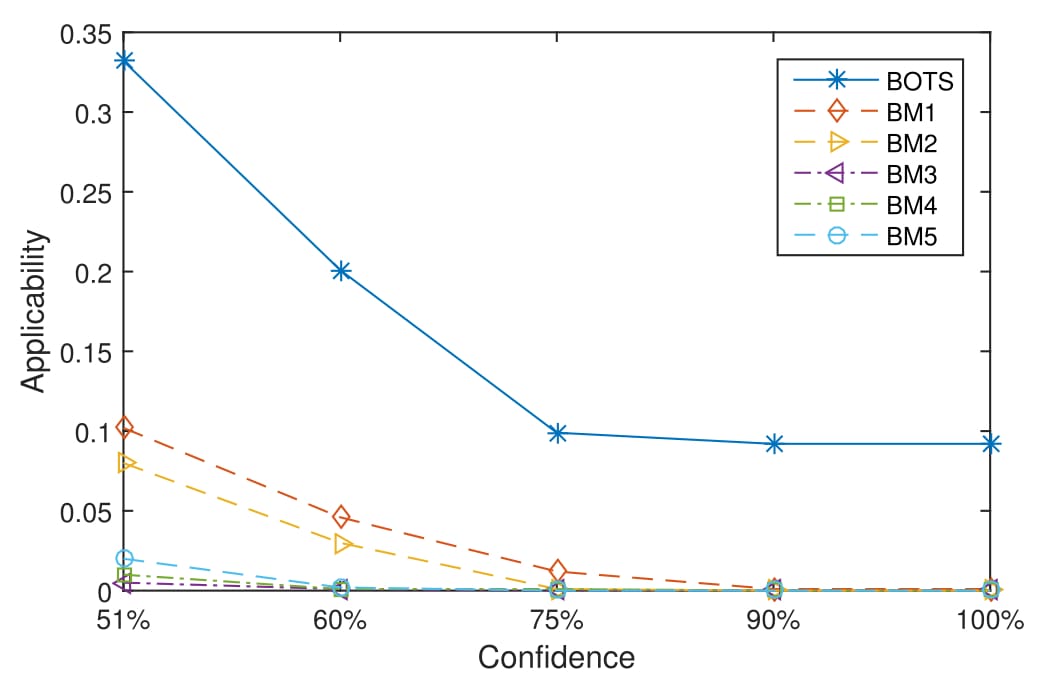}
	\caption{Applicability comparison of different segmentation approaches utilizing an individual's mobile phone data (User 10)}
	\label{fig:applicability-user10}
\end{figure} 

\begin{figure}
	\centering
	\includegraphics[width=\linewidth, keepaspectratio]{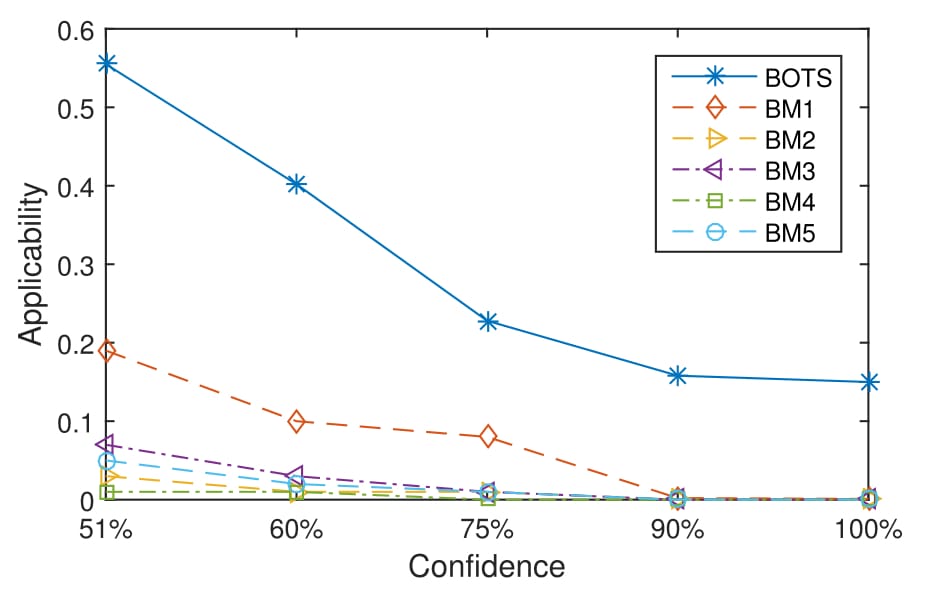}
	\caption{Applicability comparison of different segmentation approaches utilizing an individual's mobile phone data (User 51)}
	\label{fig:applicability-user51}
\end{figure} 

\begin{figure}
	\centering
	\includegraphics[width=\linewidth, keepaspectratio]{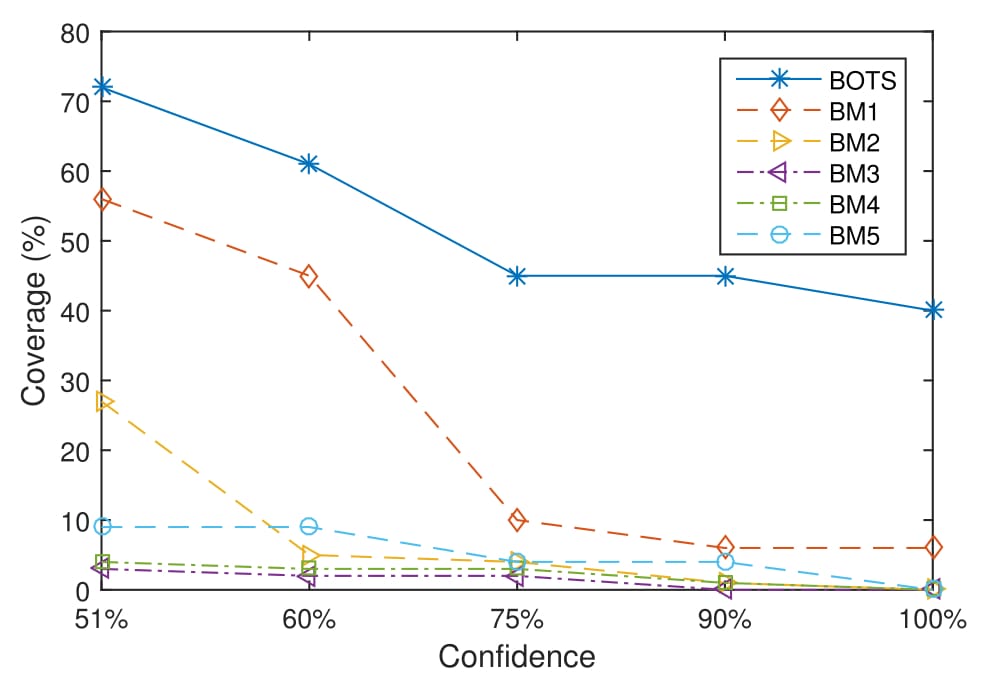}
	\caption{Data coverage (\%) comparison of different segmentation approaches utilizing an individual's mobile phone data (User 10)}
	\label{fig:coverage-user10}
\end{figure}

\begin{figure}
	\centering
	\includegraphics[width=\linewidth, keepaspectratio]{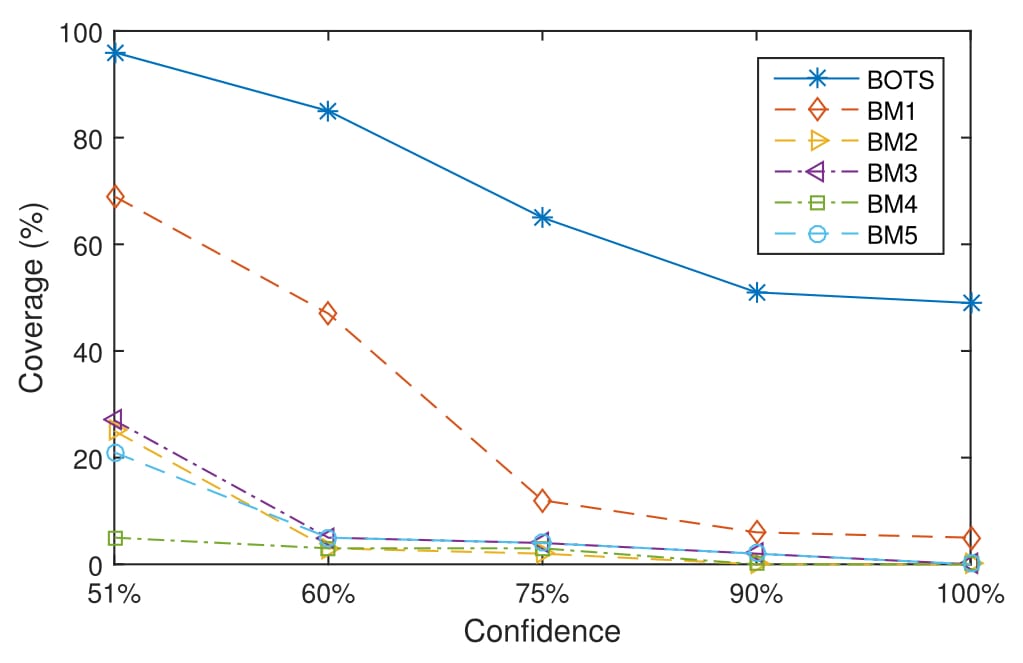}
	\caption{Data coverage (\%) comparison of different segmentation approaches utilizing an individual's mobile phone data (User 51)}
	\label{fig:coverage-user51}
\end{figure}

To show the effectiveness for individual users, Figure \ref{fig:applicability-user10} and Figure \ref{fig:applicability-user51} show the relative comparison of applicability and Figure \ref{fig:coverage-user10} and Figure \ref{fig:coverage-user51} show the relative comparison of data coverage (\%) for User 10 and User 51 respectively. For each approach, we use minimum support 1 (one instance) because no rules are meaningful below this support \cite{sarker2016behavior}. Moreover, we have explored different confidence threshold, i.e., 51\% (lowest strength), 60\% and up to 100\% (maximum strength).

From Figure \ref{fig:applicability-user10}, Figure \ref{fig:applicability-user51}, Figure \ref{fig:coverage-user10} and Figure \ref{fig:coverage-user51}, we find that our BOTS approach consistently outperforms previous approaches for different confidence thresholds. The main reason is that existing approaches do not take into account individuals' diverse behavioral patterns for segmentation in order to mine mobile user behavior. On the other-hand, our dynamic approach is individual's behavior-oriented and can capture the unique behavioral patterns for each individual user more properly, thus producing a set of behavior-oriented segments for a particular confidence threshold. 

\begin{figure}	
	\centering
	\begin{subfigure}[t]{3.9cm}
		\centering
		\includegraphics[width=3.9cm, height=2.9cm]{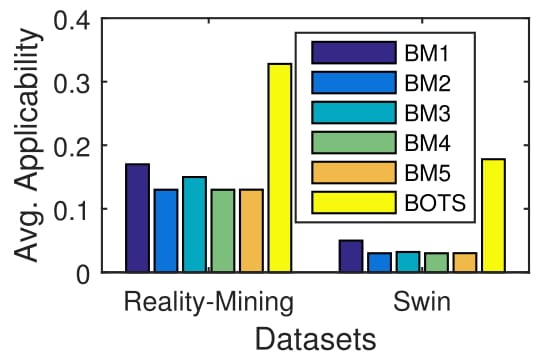}
		\caption{Average applicability}\label{fig:1a}		
	\end{subfigure}
	\quad
	\begin{subfigure}[t]{3.9cm}
		\centering
		\includegraphics[width=3.9cm, height=2.9cm]{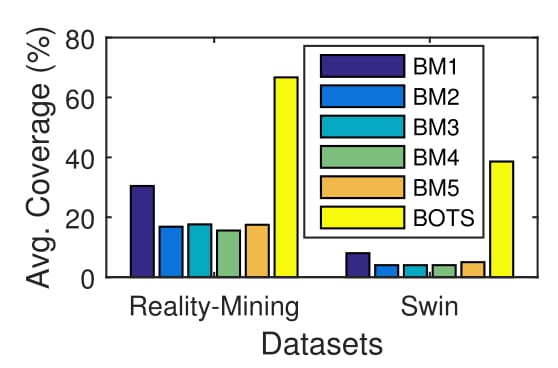}
		\caption{Average coverage}\label{fig:1b}
	\end{subfigure}
	\caption{Average applicability and average coverage comparison of different segmentation approaches utilizing the collection of individuals mobile phone data of different datasets}
	\label{fig:avg-comparison}
\end{figure}

In addition to individual's comparison, we also show the relative comparison of average applicability and data coverage (\%) for a collection of users of two different datasets shown in Figure \ref{fig:avg-comparison}. For this, we calculate the average applicability and data coverage (\%) of 30 users from reality mining dataset (randomly selected) and 15 users from swin dataset (randomly selected) for each approach with same confidence threshold 75\%. The average results also show that our BOTS approach consistently outperforms previous approaches for a collection of users. The reason is that we identify the unique behavioral patterns for each individual user more properly and get higher applicability and data coverage (\%) value for all users. However in the existing approaches, the segmentation is not individual's behavior-oriented and cannot represent the user's diverse behavioral patterns that change over-time. As a result, the possibility of masking the actual dominant behavior in a segment increases with other existing behaviors and decreases the applicability and data coverage (\%) as well for a particular confidence. In contrast, our dynamic time segmentation technique resolves these limitations and improves the segmentation quality in terms of applicability and data coverage (\%) for a particular confidence threshold by capturing individual's behavioral patterns more properly.

\subsubsection{Cross validation of Temporal Rules}
In this experiment, we show the relative comparison for prediction results of temporal rules generated using the time segments produced by different segmentation approaches utilizing individual's mobile phone data (User 10 and User 51).

\begin{figure}
	\centering
	\includegraphics[width=\linewidth, keepaspectratio]{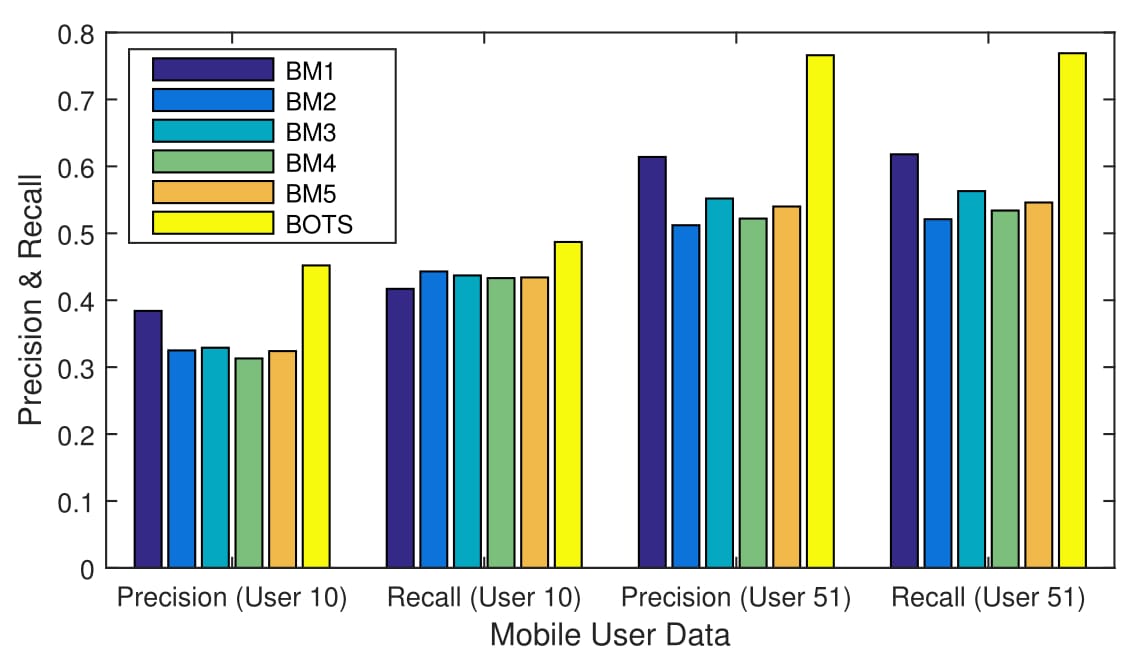}
	\caption{Precision and Recall comparison of different segmentation approaches utilizing individual's mobile phone data}
	\label{fig:nfold-individual}
\end{figure} 

As the produced rules are fully individualized, we show the prediction results in terms of precision and recall for two individuals. For this, we utilize a 10-fold cross validation on individual's mobile phone data. To be specific, we first randomly divide each dataset into ten equal parts, then we use each part as the test data while using the other parts as the training data in ten test rounds and measure the precision and recall. Figure \ref{fig:nfold-individual} shows the comparison results of different segmentation approaches for these two individuals in terms of precision and recall. 

If we observe Figure \ref{fig:nfold-individual}, we see that the produced temporal rules using our segmentation technique consistently outperforms previous approaches for different individuals, indicating that our segmentation technique produces individual's behavior-oriented segments that better capture the similar behavior of individual mobile phone users.

\section{Discussion}
\label{Discussion}
Overall, our time segmentation approach is fully individualized and behavior-oriented. Compared to the existing temporal based approaches, the applicability, data coverage (\%) and accuracy in terms of precision and recall of the discovered temporal rules are improved when our approach is used, as shown in Figures \ref{fig:applicability-user10}, \ref{fig:applicability-user51}, \ref{fig:coverage-user10}, \ref{fig:coverage-user51}, \ref{fig:avg-comparison}, \ref{fig:nfold-individual}. Among the approaches that use temporal information, our approach has the highest applicability, data coverage (\%) and accuracy, although it requires some iteration to identify the optimal base period. The following are a few key discoveries from our study. 

\begin{itemize}
	
	\item To capture the behavioral pattern of individuals, an optimal base period is the key term in our approach. However, the optimal base period can differ depending on the day of the week and from user-to-user as the behavior patterns are not identical for all individuals. In our experiments, we have discovered different base periods for different users based on different behavioral patterns.
		
		\item Another important finding of our study is that the lengths of time segments and their related support are co-related. The traditional metrics of support and confidence are not sufficient to measure the best time based rules. Thus our newly proposed applicability metric, which is the combination of temporal coverage of a segment and support value of that segment, ensures the identification of meaningful temporal segments and corresponding temporal behavior rules for a preferred confidence threshold. 
		
		\item Dynamic aggregation plays an important role for producing segments of similar dominant behavior over as long a period of time as possible over the week. The consequent time based behavior rules using these segments become more meaningful because of increased support and temporal coverage (i.e., applicability).
		
		\item We have observed a significantly lower applicability, data coverage (\%) and accuracy when using existing temporal based approaches compared to our approach. The reason is that existing approaches are not behavior-oriented and cannot capture the behavior patterns of different users' to the same degree of accuracy. Consequently, rules mined using these existing approaches have very low confidence, potentially rendering them meaningless.
		
		\item Our approach does not depend on any particular time scale, e.g., time-of-the-week, to mine individual's behavior. However, we take into account users' behaviors on a weekly basis in order to mine individual's behavior with mobile phones, as time-of-the-week is an important factor impacting on user behavior in a mobile-Internet portal and the behavior is influenced by time-of-the-week \cite{halvey2006time}. To model behavior for another time scale, e.g., time-of-the-day, day-of-month, week-of-month, week-of-year or quarter-of-year, corresponding data pre-processing is needed according to these scales before applying the segmentation approach.	
\end{itemize}

\section{Applications of BOTS}
\label{Applications of BOTS}
As we produce behavior-oriented time segments according to individual's behavioral patterns, these segments can be used in various real-life applications to assist them intelligently. Hence, we summarize some real-life applications related to temporal segments and corresponding mobile phone usages behavior of individuals.	These are:

\subsection{Call Firewall}
Call firewall basically monitors and handles incoming calls by keeping unsolicited and unwanted
calls away while allowing desired calls to pass through. Unlike email spam, call spam is a real-time problem which requires a real-time defense mechanism \cite{phithakkitnukoon2011behavior}. The real challenge is thus to block the spam call before the phone rings. Not only do these spam calls create a nuisance for the user, each incoming phone call creates different levels of nuisance depending on the user's present mood or state of mind based on situational, spatial, and temporal contexts \cite{kolan2008nuisance}. Therefore, a set of temporal firewall rules can be discovered using our BOTS approach, e.g., IF calls come between 10:00AM-11:00AM, THEN forward it to voicemail, IF calls come between 4:45PM-5:30PM, THEN drop the call.

\subsection{Planning and Scheduling}
Predicting incoming calls can be very useful for planning and scheduling \cite{phithakkitnukoon2011towards} like weather forecasting. People normally check weather forecast before leaving homes and watch for signs of approaching storms to prepare and schedule their days accordingly. Knowing what is coming next gives us supplemental time to think, prepare, and optimize our solutions. Therefore, we believe that incoming call prediction based on temporal information can also be useful for daily planning and it may become an important element as an initiative decision support for our daily life scheduling.

\subsection{Phone Call Interruption Management}
Mobile phones are considered to be `always on, always connected' device but the mobile users are not always attentive and responsive to incoming communication \cite{chang2015investigating}. For this reason, sometimes people are often interrupted by incoming phone calls which not only disturb the phone users but also can disturb the people nearby. Such kind of interruptions may create embarrassing situation not only in an official environment, e.g., meeting, lecture etc. but also affect in other activities like examining patients by a doctor or driving a vehicle etc. Sometimes these kind of interruptions may reduce worker performance, increased errors and stress in a working environment \cite{pejovic2014interruptme}. Therefore, in order to minimize such interruptions, individual's phone call response behavior-oriented time segments can be used to build intelligent call interruption management system.

\subsection{Phone Call Reminder}
One of the common problems of everyday life is forgetting to make a phone call that could either be an event-based call such as birthday call, meeting planning call, etc., or a nonevent-based call such as calling parents on weekends, calling girlfriend/boyfriend during a lunch break, etc. \cite{phithakkitnukoon2011behavior}. Therefore, the outgoing phone call behavioral time segments discovered by our BOTS approach can help to generate a ``reminder'' for the user to place a call to a particular person based on the user's past calling history.

\subsection{Enhancing Phone Usability}
Predicting outgoing calls can be useful for enhancing mobile phone's usability by providing a list of the most likely contacts/numbers to be dialed when the user wants to make a call \cite{phithakkitnukoon2011towards}. Therefore, the outgoing phone call behavioral time segments discovered by our BOTS approach can help to reduce the searching time as well as enable better life synchronization for the users.

\subsection{Mobile Phone Notification Management}
Mobile phone notifications are increasingly used by a variety of applications to inform users about events, news or just to send alerts and reminders to them \cite{mehrotra2016prefminer}. However, many notifications are neither useful nor relevant to the users' interests and, also for this reason, they are considered disruptive and potentially annoying. Some examples of such notifications are promotional emails, game invites on social networks and predictive suggestions by applications, e.g., Twitter, Facebook, WhatsApp. According to \cite{mehrotra2016prefminer}, users mostly dismiss (i.e., swipe away without clicking) notifications that are not useful or relevant to their interests.  Therefore, in order to minimize such interruptions, individual's interaction rules with their mobile phones based on time can be used to build intelligent mobile phone notification management system.

\subsection{Personalized Apps Recommendation}
With the rapid development and adoption of mobile platforms such as smartphones and tablets, they have become one of the most important media for social entertainment and information acquisition \cite{zhu2014mining}. In fact, the temporal context and corresponding app usages (e.g., Multimedia, Facebook, Gmail, Youtube, Skype, Game) data is recorded in context-rich device logs which can be used for mining the personal context-aware preferences of mobile phone users that is, which app is preferred by a particular user under a certain context. Particularly, mining such preferences is a fundamental work for understanding the app usages behaviors of mobile phone users. Therefore, the extracted temporal behavior rules utilizing context-logs can be used to provide personalized context-aware recommendation of different mobile phone apps (e.g., Multimedia, Facebook, Gmail, Youtube, Skype, Game) for the mobile phone users.

\section{Conclusion and Future Work}
\label{Conclusion and Future Work}
In this paper, we have introduced a dynamic behavior-oriented time segmentation approach for extracting temporal behavior rules, in order to mine mobile user behavior utilizing their mobile phone data. Our approach dynamically identifies the optimal continuous time segments, each of which is dominated by a particular behavior of the user. Consequently, temporal rules are formulated for these time segments, which can be used for developing an automated rule-based personal assistance system for mobile phone users. The time segments are identified based on the contiguous dominant behavior of the users, can have different spans over the week, and will be different from user-to-user to truly reflect their behavioral patterns. Furthermore, the time segments and corresponding behavioral rules are determined in such a way that maximum temporal coverage by the rules is achieved for the preferred confidence threshold, to achieve maximum applicability for the rules. For this purpose, we have also introduced the applicability measure, which takes into account the support and temporal coverage that the mined rules achieve. Our experiments on real life datasets have shown that individuals do have different time segmentations and related behaviors. Although we choose phone call behavior contexts as examples, our approach is also applicable to other application domains. We believe that our approach opens a promising path for future research on extracting behavioral rules of individuals based on time-series data. 

In future work, we plan to enlarge our behavior mining problem by incorporating additional contexts such as location, social relationship between individuals, and social situation, in order to discover behavioral rules for individual mobile phone users based on multi-dimensional contexts.

\bibliographystyle{compj}
\bibliography{segmentation-bibfile}

\end{document}